\pdfoutput=1

\documentclass[useAMS,usenatbib]{mnras}
\usepackage{graphicx}
\usepackage{natbib}
\usepackage{xcolor}
\usepackage{longtable,lscape}

\def\kms{$\mbox{km s}^{-1}$}

\newcommand{\msun}{{\rm M}_\odot}
\newcommand{\dr}{{\rm d}}

\def\aap{A\&A}
\def\aapr{A\&A Rev.}
\def\apj{ApJS}
\def\apjl{ApJL}
\def\apjs{ApJ Supp.}

\def\mnras{MNRAS}
\def\aj{AJ}
\def\nat{Nature}
\def\aaps{A\&AS}

\def\pasp{PASP}

\usepackage[normalem]{ulem}

\usepackage{txfonts}

\title[On the black hole content and IMF of 47 Tuc]{On the black hole content and initial mass function of 47 Tuc}
\author[V. H\'{e}nault-Brunet et al.]
{V. H\'{e}nault-Brunet$^{1,2}$\thanks{E-mail: vincent.henault@smu.ca}, M. Gieles$^{3,4,5}$, J. Strader$^{6}$, M. Peuten$^{5}$, E. Balbinot$^{7,5}$, \newauthor K.E.K. Douglas$^{8}$\\
$^1$Department of Astronomy and Physics, Saint Mary's University, 923 Robie Street, Halifax, NS B3H 3C3, Canada\\
$^2$National Research Council, Herzberg Astronomy \& Astrophysics, 5071 West Saanich Road, Victoria, BC, V9E 2E7, Canada\\
$^3$Institut de Ci\`{e}ncies del Cosmos (ICCUB), Universitat de Barcelona, Mart\'{i} i Franqu\`{e}s 1, 08028 Barcelona, Spain\\
$^4$ICREA, Pg. Lluis Companys 23, 08010 Barcelona, Spain.\\
$^5${Department of Physics, University of Surrey, Guildford, GU2 7XH, Surrey, UK}\\
$^6${Department of Physics and Astronomy, Michigan State University, East Lansing, MI 48824, USA} \\
$^7$Kapteyn Astronomical Institute, University of Groningen, Postbus 800, NL-9700AV Groningen, the Netherlands  \\
$^8$Department of Physics and Astronomy, University of Victoria, PO Box 3055, STN CSC, Victoria BC V8W 3P6, Canada
}

\hypersetup{draft}
\begin{document}

\date{Accepted 2019 October 19. Received 2019 October 18; in original form 2019 August 20}

\pagerange{\pageref{firstpage}--\pageref{lastpage}} \pubyear{2019}

\maketitle

\label{firstpage}

\begin{abstract}
The globular cluster (GC) 47 Tuc has recently been proposed to host an intermediate-mass black hole (IMBH) or a population of stellar-mass black holes (BHs). To shed light on its dark content, we present an application of self-consistent multimass models with a varying mass function and content of stellar remnants, which we fit to various observational constraints. Our best-fitting model successfully matches the observables and correctly predicts the radial distribution of millisecond pulsars and their gravitational accelerations inferred from long-term timing observations. The data favours a population of BHs with a total mass of $430^{+386}_{-301}$ $\msun$, but the most likely model has very few BHs. Since our models do not include a central IMBH { and accurately reproduce the observations}, we conclude that there is currently no need to invoke the presence of an IMBH in 47 Tuc. The global present-day mass function inferred is significantly depleted in low-mass stars (power-law slope $\alpha=-0.52^{+0.17}_{-0.16}$). Given the orbit and predicted mass-loss history of this massive GC, the dearth of low-mass stars is difficult to explain with a standard initial mass function (IMF) followed by {long-term preferential escape of low-mass stars} {driven by two-body relaxation}, and instead suggests that 47 Tuc may have formed with a bottom-light IMF. { We discuss alternative evolutionary origins for the flat mass function and ways to reconcile this with the low BH retention fraction.} Finally, by capturing the effect of dark remnants, our method offers a new way to probe the IMF in a GC above the current main-sequence turn-off mass, for which we find a slope of $-2.49\pm0.08$.

\end{abstract}

\begin{keywords}
galaxies: star clusters -- globular clusters: general -- globular clusters: individual: 47 Tuc -- stars: kinematics and dynamics -- stars: luminosity function, mass function -- stars: black holes.
\end{keywords}

\section{Introduction}

\label{Section_intro}

Postulated to populate the mass regime between stellar-mass black holes (BHs) and supermassive black holes (SMBHs), intermediate-mass black holes (IMBHs; $\sim10^2-10^5 \ \msun$) could represent a missing link in black hole growth, seeding the rapid formation of SMBHs in galaxies in the early Universe \citep[e.g.][]{2010A&ARv..18..279V}. Theoretical studies have suggested that IMBHs may form in the dense environment found in the cores of globular clusters (GCs), for example { from a supermassive star that forms} via runaway stellar collisions during their early life (e.g. \citealt{2004Natur.428..724P}; \citealt{2018MNRAS.478.2461G}) or { via} mergers of BHs via gravitational wave inspiral over a period of Gyrs (\citealt{2015MNRAS.454.3150G}; \citealt*{2019MNRAS.486.5008A}). From simply extrapolating the observed relation between the black hole mass and the velocity dispersion of its host galaxy \citep{2000ApJ...539L...9F,2000ApJ...539L..13G}, one may expect black holes with masses characteristic of IMBHs to hide in GCs. If confirmed, observational evidence of IMBHs would also have a significant impact on a wide range of other astrophysical problems, from the origin of ultraluminous X-ray sources in nearby galaxies to the rates of gravitational wave events.

The quest to find IMBHs in the core of Milky Way GCs, where their presence has been debated for a long time, is however challenging. Deep radio observations of nearby GCs looking for signatures of accretion onto an IMBH have put upper limits on their masses of a several 100 to $\sim1000$ $\msun$ \citep{2012ApJ...750L..27S, 2018arXiv180600259T}, but these non-detections could be due to the paucity of gas in GCs at the present day.  A lot of effort has thus focused on indirect dynamical inference to search for the presence of IMBHs, but to this day all reported detections remain contentious and inconclusive. 

Earlier claims of IMBH detections in GCs are based on indications of a shallow central cusp in the surface brightness profile or a central rise in the velocity dispersion profile inside the sphere of influence of a putative IMBH \citep[e.g.][]{1976ApJ...208L..55N, 2008ApJ...676.1008N, 2011A&A...533A..36L}, which is expected from relaxation arguments \citep{1976ApJ...209..214B}. 
All these findings have been rebutted, or at least questioned. In some cases, this was due to improvements in the measurements of the central velocity dispersion with discrete kinematics \citep[proper motions or line-of-sight velocities for individual stars;][]{2010ApJ...710.1063V, 2013ApJ...769..107L} rather than integrated-light spectra which can be biased by stochasticity induced by bright stars \citep[see e.g.][]{2015MNRAS.453..365B, 2017MNRAS.467.4057D}. In other cases, there are other physical explanations that cannot be ruled out and are more plausible. Radially-biased anisotropy in the velocity distribution \citep*{2017MNRAS.468.4429Z} or the presence of dark remnants such as white dwarfs, neutron stars \citep{1977ApJ...218L.109I,  2014MNRAS.438..487D}
 and BHs (\citealt*{2013A&A...558A.117L}; \citealt{2016MNRAS.462.2333P}; \citealt*{2019MNRAS.482.4713Z}; \citealt{ Baumgardt2019}) can mimic the dynamical signatures previously ascribed to an IMBH, in particular the increase in the central velocity dispersion. The same goes for high-velocity stars found in the cores of some GCs (\citealt*{1991ApJ...383..587M, 1979AJ.....84..752G}; \citealt{2012A&A...543A..82L}), which may suggest interaction with an IMBH but can in fact simply result from interactions with a binary system \citep{2012A&A...543A..82L} or represent `potential escapers', stars above the local escape velocity that are energetically unbound but trapped inside the Jacobi surface for several orbits before escaping the cluster \citep*{2000MNRAS.318..753F,2017MNRAS.466.3937C,2017MNRAS.468.1453D}.

Pulsar timing is a promising way to probe the gravitational potential of GCs and look for an IMBH in their core. Given the precise and stable clock provided by measurements of the period (spin or { binary} orbital period) of a pulsar, one can relate an observed change in that period (typically over a baseline of several years) to the gravitational acceleration from the cluster potential at the position of the pulsar. If there is no process intrinsic to the pulsar or binary system changing this period, then the observed period derivative is directly related to the gravitational acceleration along the line-of-sight \citep{1993ASPC...50..141P, 2017ApJ...845..148P}. \citet{2017MNRAS.468.2114P} recently used long-term timing observations of the millisecond pulsar (MSP) PSR B1820$-$30A, located less than $0.5^{\prime\prime}$ from the centre of NGC 6624, to argue for the presence of a central IMBH in that GC. \citet{2018MNRAS.473.4832G} however showed that the maximum line-of-sight acceleration at the position of the MSP is consistent with a mass segregated dynamical model of NGC 6624 without an IMBH. \citet{Baumgardt2019} also concluded that there is no need for an IMBH in NGC 6624 by comparing observations to $N$-body models with and without an IMBH.

Using spin period derivatives for 23 MSPs with timing solutions in 47 Tuc, \citet*{2017Natur.542..203K} argued for an IMBH with a mass $M_{\bullet}\sim 2200^{+1500}_{-800} \ \msun$ in this cluster. To reach this conclusion, they compared the inferred pulsar accelerations (assuming an intrinsic spin-down distribution comparable to that of the Galactic field MSP population) to the accelerations of neutron stars in a grid of $N$-body simulations of GCs with and without an IMBH. Several assumptions could however have affected their analysis, for example the short distance to 47 Tuc of 4\,kpc assumed by \citet{2017Natur.542..203K}, whereas most recent estimates put it at $\sim4.5$\,kpc \citep[][and references therein]{2018ApJ...867..132C}.
As pointed out by \citet{2019ApJ...875....1M}, the $N$-body models of \citet{2017Natur.542..203K} do not have significant mass in binaries or other heavy objects (their main grid of models contained no primordial binaries and little or no stellar-mass black holes after a Hubble time). A lack of massive objects that would sink to the centre due to dynamical friction could lead to an underestimation of the gravitational acceleration in the central regions of their models without an IMBH. The assumption by \citet{2017Natur.542..203K} of a mass 1.4 $\msun$ for the MSPs \citep[a lower limit on the true masses given that many of these are in binary systems; e.g.][]{2017MNRAS.471..857F} could also bias the comparison of the inferred accelerations of some pulsars with the distribution of accelerations in the $N$-body models since the adopted mass affects the predicted spatial distribution of these objects. \citet{2019ApJ...875....1M} used stellar proper motions from the {\it Hubble Space Telescope} (HST) in the core of 47~Tuc along with Jeans models to conclude that the measured central velocity dispersion profile provides no strong evidence for an IMBH, in particular if significant concentrated populations of heavy binary systems and dark stellar remnants are included.

This last point is particularly relevant given the recent interest in uncovering BH populations in GCs because of the detection of gravitational waves from merging BHs by LIGO \citep{2016PhRvL.116f1102A}. Dynamical formation of close BH-BH binaries in the dense cores of GCs could be one of the main formation channels for these mergers \citep[e.g.][]{1993Natur.364..423S,2000ApJ...528L..17P,2016ApJ...818L..22A, 2016PhRvD..93h4029R}. The amount of dynamically formed BH-BH binaries and eventual mergers in GCs as a function of time depends on several ingredients (e.g. the initial mass function -- IMF, the initial-final mass relation, massive binary star evolution, the magnitude of natal kicks that BHs receive at birth, the dynamical age of the cluster, etc.), many of which are poorly constrained. A promising avenue to make progress is to quantify the size of BH populations in GCs today.

Radio and X-ray observations of GCs have led to the discovery of a few accreting stellar-mass BH candidates in mass-transferring binaries \citep[e.g.][]{2012Natur.490...71S}, including one in 47 Tuc \citep{2015MNRAS.453.3918M}. \citet{2018MNRAS.475L..15G} identified a detached stellar-mass BH candidate in the core of the Galactic GC NGC~3201 from the radial velocity variations of a main-sequence turn-off star companion. These findings suggest that the natal kick velocities of BHs can be low enough to retain at least a small fraction of the BHs formed in GCs. Since these types of binaries are rare \citep{2018ApJ...852...29K}, they could represent the tip of the iceberg of a much larger population of BHs that might be common in GCs\footnote{Note that models from \citet{2018ApJ...852...29K} show little correlation between the presence of mass-transferring BH systems and the total number of BHs in the cluster at any given time; the net rate of formation of BH-non-BH binaries is largely independent of the total number of retained BHs. Should this lack of correlation generally hold true, one would not be able to estimate the total number of BHs in a GC by extrapolating from the number of known BH candidates in binaries.} but would have to be detected indirectly (for example through their dynamical signature).

Due to two-body relaxation and the trend towards kinetic energy equipartition, BHs (which are $\gtrsim10$ times more massive than the typical $\sim0.5 \ \msun$ star in a GC) have a dynamical heating effect on cluster stars and tend to inflate the visible cluster core while suppressing mass segregation among observable stars in the cluster \citep[e.g.][]{2004ApJ...608L..25M, 2008MNRAS.386...65M, 2016MNRAS.462.2333P}. Based on relations between the number of BHs in GC models and their global properties or kinematics, present-day populations of a few tens to a few hundreds of BHs have been proposed in several Milky Way GCs \citep[e.g.][]{2016MNRAS.462.2333P, 2018ApJ...855L..15K, 2019ApJ...871...38K, 2018MNRAS.478.1844A}. For 47 Tuc, \citet{2018ApJ...864...13W} inferred a population of $\sim20$ BHs (but possibly as large as $\sim150$ BHs within $2\sigma$) based on the observed mass segregation between giants and low-mass main-sequence stars in the central regions of the cluster. There is however a need to confirm the predictions from these simple relations with more elaborate methods and/or by comparing models simultaneously to a wider range of observations of the internal kinematics and structural properties of GCs.

To shed light on the mass distribution and dark content  of 47~Tuc, we present in this work a new self-consistent equilibrium multimass model of the cluster (without an IMBH) that captures the effect of mass segregation in the presence of dark stellar remnants. In Section~\ref{data_section}, we present the various datasets to which we fit our models. We describe the multimass models and fitting procedure in Section~\ref{multimass_models_section} and the results of the fits in Section~\ref{results_section}.  We discuss the implications for the BH content, global present-day mass function and IMF of 47 Tuc in Section~\ref{discussion_section}, and finally we summarize our conclusions in Section~\ref{conclusion_section}.

\section{Data}
\label{data_section}

We summarize below the various datasets to which the models described in Section \ref{multimass_models_section} are fitted, as well as additional data to which we compare predictions from our best-fitting models. In contrast to previous work that studied the mass distribution and black hole content of 47~Tuc focusing on one type of observable \citep[e.g. either pulsar accelerations, the central velocity dispersion, or the observed mass segregation;][see Section~\ref{Section_intro}]{2017Natur.542..203K, 2019ApJ...875....1M, 2018ApJ...864...13W}, we simultaneously compare and fit our models to several observables that probe the mass distribution within the cluster and the phase-space distribution of objects of different masses.

\subsection{Number density profile}
\label{ND}

To constrain the structural parameters of the cluster, we use the number density profile of 47~Tuc from the catalogue of \citet{2019MNRAS.tmp..651D}, which is based on {\it Gaia} Data Release 2 (DR2) data, allowing accurate membership selection in the outer parts of the cluster. The profile is extended using supplemental literature data from surface brightness measurements \citep{1995AJ....109..218T} in the central regions of the cluster, where the completeness of {\it Gaia} data is affected by crowding. The {\it Gaia} number density profile is calculated from bright stars ($m>0.6 \ \msun$) and the literature data are also dominated by bright stars. In both cases, the profile is dominated by stars from a narrow range of stellar masses \citep[cf.][]{2019MNRAS.tmp..651D}. When fitting models, we thus assume that the number density profile traces the spatial distribution of upper main-sequence and evolved stars.

\subsection{Kinematics}

\subsubsection{Proper motions}
\label{PMs}

As the main constraint on the total mass of the cluster (and its distribution as a function of distance from the centre), we consider kinematic data from various sources. We use the {\it HST} proper motion dispersion profiles (tangential and radial components in the plane of the sky) in the central regions of the cluster from \citet{2015ApJ...812..149W}. These are based on cleaned samples of bright stars from the {\it HST} proper motion catalogs of \citet{2014ApJ...797..115B} and only include stars brighter than the main-sequence turn-off.

We complement the dataset above with {\it HST} proper motion dispersion profiles of main-sequence stars from \citet{2017ApJ...850..186H} in a field located further out and centered 6.7$^\prime$ in projection from the cluster centre \citep[$\sim8.8$\,pc, a little over two times the half-light radius;][]{2005ApJS..161..304M}. The mean mass of the stars in this dataset is 0.38 $\msun$ - significantly lower than the stars entering the other kinematic datasets - so inclusion of this dataset provides constraints on mass segregation and the mass dependence of kinematics. These data show evidence of radial anisotropy in the outer parts of 47 Tuc and provide a useful constraint on the anisotropy parameter (see Section \ref{multimass_models_section}). As shown by \citet{2017MNRAS.468.4429Z}, anisotropy in the outer regions of a cluster can influence the central velocity dispersion, with radially anisotropic models having a larger central dispersion (in the radial, tangential, and line-of-sight components) than the equivalent isotropic models. This ingredient was however not considered by \citet{2019ApJ...875....1M} who fitted isotropic Jeans models to the velocity dispersion profile in the core of the cluster. Given the different mass regime probed by these data compared to the other kinematic data used, they also provide additional leverage to capture and correctly model the effect of mass segregation. 

 For our main set of models, we adopt a distance of 4.45 kpc to convert model velocities to observed proper motions in units of mas~yr$^{-1}$. This distance is the most recent and precise estimate ($4.45\pm0.01\pm0.12$ kpc; random and systematic errors, respectively) 
for 47 Tuc, obtained by \citet{2018ApJ...867..132C} using parallaxes from {\it Gaia} DR2 and taking advantage of the background stars in the Small Magellanic Cloud and quasars to account for parallax systematics. This distance also agrees with the recent dynamical distance of $4.4\pm0.1~$kpc by \citet{2018MNRAS.473.5591K}. In Section \ref{results_section}, we will discuss how 2$\sigma$ excursions from this adopted value of $D=4.45$~kpc influence our results, and show that our main conclusions remain unaffected even with $D=4.2$~kpc or $D=4.7$~kpc.

\subsubsection{Line-of-sight velocities}
\label{sigma-LOS}

We also use the line-of-sight velocity dispersion profile of 47 Tuc from \citet{2018MNRAS.478.1520B}, which builds on a compilation of line-of-sight velocity measurements from \citet{2017MNRAS.464.2174B} and considers additional archival data. This includes several measurements from the literature \citep{1983A&AS...54..495M, 1995AJ....110.1699G, 2009A&A...505..117C, 2011A&A...530A..31L, 2013A&A...549A..41G, 2016MNRAS.455..199D, 2017AJ....153...75K, 2018MNRAS.473.5591K} 
which have been put on the same velocity scale by matching their mean radial velocity.

\subsection{Stellar mass functions}

\label{MF}

In order to constrain the global stellar mass function of the cluster, we compare our models to measurements of the local stellar mass function in concentric annuli covering different ranges of projected distances from the cluster centre. \citet{2017MNRAS.471.3668S} extracted stellar mass functions in four annuli within a projected radius of $1.6^{\prime}$ from the centre of 35 GCs using the {\it HST}/ACS photometry published by the Globular Cluster ACS Treasury Project \citep{2007AJ....133.1658S}. To convert magnitudes to stellar masses, they adopted
the mass-luminosity relation of suitable isochrones from the \citet{2007AJ....134..376D} database. 47 Tuc was not retained in the final sample analysed in \citet{2017MNRAS.471.3668S}, but stellar mass functions determined in the same way as for their 35 selected clusters were provided to us by A. Sollima (priv. comm.). These stellar mass functions for 47 Tuc are also presented in figure 2 of \citet{2018MNRAS.478.1520B}. The annuli cover the projected radius ranges $0.0^{\prime}< R < 0.4^{\prime}$, $0.4^{\prime}< R < 0.8^{\prime}$, $0.8^{\prime}< R < 1.2^{\prime}$, and $1.2^{\prime}< R < 1.6^{\prime}$. The mass range covered is $\sim0.2 \ \msun < m < 0.85 \ \msun$ in the outermost of these four annuli but is narrower in the innermost annulus where crowding significantly affects the completeness of observations for the fainter low-mass stars.

Note that observations of the mass function in 47 Tuc were not considered in the analysis of \citet{2017Natur.542..203K}, who argued in favour of an IMBH. These authors used $N$-body simulations of isolated star clusters with a \citet{2001MNRAS.322..231K} IMF (and hence no significant preferential loss of low-mass stars), despite the fact that previous work suggests a global stellar mass function depleted in low-mass stars for this cluster \citep[e.g.][]{2011MNRAS.410.2698G}. This could have affected their inferred mass distribution, and hence the predicted pulsar accelerations on which their conclusions are based.

\subsection{Pulsar data}
\label{pulsar_data}

Although we do not directly include them in the fitting procedure, we consider the spatial distribution and inferred accelerations of pulsars in 47 Tuc as additional observables that are compared to our best-fitting models to serve as a consistency check. Because pulsars have a higher mass than our tracer stars and their accelerations probe the gravitational potential of the cluster, these data help to assess the validity and predictive power of our models fitted to the number density profile, kinematics, and stellar mass function data.

\citet{2017MNRAS.471..857F} presented updated or new long-term timing solutions for the majority of the 25 known millisecond pulsars in 47 Tuc based on data from the Parkes 64-m radio telescope. To compare with the maximum line-of-sight accelerations predicted by our models, we take the orbital period derivatives and associated inferred line-of-sight accelerations ($a_{\rm los} \simeq c \ \dot{P}_{\rm orb}/P_{\rm orb}$, i.e. no intrinsic component to the observed period derivative) for 10 binary systems with a timing solution and measured orbital period derivative (\citealt{2016MNRAS.462.2918R} for 47 Tuc X, \citealt{2017MNRAS.471..857F} for the nine other binary systems: 47 Tuc E, H, I, Q, R, S, T, U, and Y). Two of these (47 Tuc I and R) are black-widow systems with a low-mass companion ($<0.05 \ \msun$) and the others have a white dwarf companion. For the remainder of the MSPs with a timing solution (either isolated pulsars or binaries for which the orbital period derivative could not be measured), we compute an upper limit on the line-of-sight acceleration from the measured first spin period derivative ($a_{\rm los} \leq c \ \dot{P}_{\rm spin}/P_{\rm spin}$; with measurements from \citealt{2017MNRAS.471..857F}, \citealt{2016MNRAS.462.2918R}, and \citealt{2018MNRAS.476.4794F}).

From the coordinates of all of the above MSPs, plus two additional ones without a timing solution (47 Tuc P and V; e.g. \citealt{2016MNRAS.462.2918R}), we build a cumulative radial distribution of MSPs in 47 Tuc which we use as a proxy for the spatial distribution of neutron stars in the cluster and an additional sanity check for our mass segregated models. We adopt the position of the cluster centre determined by \citet{2010AJ....140.1830G}.

\section{Models and fitting procedure}

\label{multimass_models_section}

\subsection{Multimass dynamical models}

To model the mass distribution within 47 Tuc, we use the multimass version of the {\sc limepy}\footnote{A {\sc python} solver allowing to compute models is available at \url{https://github.com/mgieles/limepy}} family of dynamical models \citep{2015MNRAS.454..576G}. These models are based on a distribution function that approximates an isothermal sphere for the most bound stars in the central regions and is described by polytropes in the external regions near the escape energy. The latter aspect allows to describe dynamically evolved and tidally limited systems and  approximately capture the effect of the galactic tidal field on the structure and dynamics of GCs in the simplifying assumption of spherical symmetry. The truncation parameter $g$ (related to the polytropic index $n = g + 1.5$) sets the sharpness of the truncation in energy: models with larger values of $g$ are more extended and have a less abrupt truncation. The concentration of the models is controlled by the dimensionless central potential $W_0$, similar to the concentration parameter of \citet{1966AJ.....71...64K} models. {\sc limepy} models with $g=1$ are single-mass King models, while models with $g=0$ are \citet{1954MNRAS.114..191W} models and those with $g=2$ the also well-known isotropic and non-rotating \citet{1975AJ.....80..175W} models. Note that any intermediate level of truncation `sharpness' is possible by setting $g$ to a non-integer value. The exact meaning of $W_0$ depends on the definition of the mean mass, and we adopt here the global mean mass of the entire model \citep[for details on this and alternative definitions, see][]{2017MNRAS.470.2736P}. The distribution function of the {\sc limepy} models includes an angular momentum term that allows to include Michie-type anisotropy \citep[isotropy in the centre of the cluster, radially anisotropic velocity distribution in the intermediate parts, and isotropy again near the truncation radius;][]{1963MNRAS.125..127M}. This is specified by the anisotropy radius ($r_{\rm a}$) parameter, which sets the amount of velocity anisotropy in the system: models with a small $r_{\rm a}$ are radially anisotropic, while models with a large $r_{\rm a}$ (with respect to the truncation radius) have an isotropic velocity distribution everywhere. This form of anisotropy profile has been shown to satisfyingly describe globular clusters evolving in an external tidal field in $N$-body simulations \citep{2015MNRAS.451.2185S,2016MNRAS.462..696Z, 2016MNRAS.455.3693T}. \citet{2017MNRAS.470.2736P} showed with $N$-body models that in the early evolution $r_{\rm a}$ is smaller for the lower-mass objects and in the late stages of evolution it is the more massive stars and remnants that have a smallest $r_{\rm a}$. For the present-day dynamical age of 47 Tuc \citep{2011MNRAS.410.2698G}, the results of \citet{2017MNRAS.470.2736P} suggest that $r_{\rm a}$ is approximately independent of mass, hence we proceed with the assumption that anisotropy for all mass bins can be described by a single value of $r_{\rm a}$.

Multiple mass components and the effect of mass segregation between these components are included in the {\sc limepy} models by relating the velocity scale of each mass component ($s_j$) to the mass of the component ($m_j$) as $s_j \propto m_j^{-\delta}$ \citep[where $\delta$ is usually set to 1/2; see][]{1979AJ.....84..752G}. This formulation leads to equipartition at high masses, but to a shallower mass dependence of the central velocity dispersion at lower masses which matches what is seen in $N$-body and Monte Carlo simulations of GCs \citep[e.g.][]{2015MNRAS.454..576G, 2016MNRAS.458.3644B, 2017MNRAS.470.2736P}. We emphasize that the {\sc limepy} models have been extensively tested against snapshots from direct $N$-body simulations for which the full phase-space information is available, including that of different mass components in simulations with a mass spectrum \citep{2016MNRAS.462..696Z, 2017MNRAS.470.2736P}. In particular, \citet{2017MNRAS.470.2736P} showed that multimass {\sc limepy} models accurately reproduce the degree of mass segregation in evolved multimass systems in different tidal fields and with vastly different populations of dark remnants (e.g. from no neutron stars and BHs to a large retention fraction of stellar-mass black holes). 

In addition to the three parameters introduced above ($W_0$, $g$ and $r_{\rm a}$) and the mean stellar mass ($m_j$) and total mass ($M_j$) of each of the different mass components, two physical scales must be specified to compute a model that can be compared to observations. We adopt here the total cluster mass ($M$) and half-mass radius ($r_{\rm h}$) as the additional parameters specifying those scales.

\subsection{Stellar mass function and remnants}

To determine the individual values of $m_j$ and $M_j$, we define four additional free parameters associated with the global mass function within the cluster. An initial mass function is adopted and defined by a three-component broken power law ($\dr N/\dr m \propto m^{-\alpha}$) with break masses at $m=0.5 \ \msun$ and $m=1 \ \msun$, and power-law indices $\alpha_1$, $\alpha_2$ and $\alpha_3$ corresponding to the low, intermediate, and high-mass ranges, respectively. We evolve this mass function to the present day assuming a metallicity of ${\rm [Fe/H]}=-0.7$ and an age of 11 Gyr \citep{2010ApJ...708..698D, 2013Natur.500...51H} to turn higher mass stars into white dwarfs, neutron stars, and black holes following the mass function evolution algorithm from \citet{TheBalb17}, updated\footnote{\url{https://github.com/balbinot/ssptools/blob/master/ssptools/evolve_mf.py}} by Peuten et al. (in prep.) to include a more realistic treatment of the initial-final mass relation for the different types of remnants \citep[e.g.][]{2016ApJS..222....8D, 2008ApJS..174..223B}. Modification of the mass function by dynamical evolution and preferential escape of low-mass stars and remnants is assumed to be negligible for 47 Tuc (see Section \ref{IMF}), so this effect is not included in our models and fitting procedure. { We note that Monte Carlo models by \citet{2011MNRAS.410.2698G} suggest that almost 50\% of the mass of the cluster has been lost, but this is due to stellar evolution mass loss, i.e. high mass stars turning into remnants, and not  escaping low-mass stars.}

We allow for a varying retention fraction of the number of BHs ($BH_{\rm ret}$), removing the more massive BHs first as expected for a slow depletion of the BH reservoir from dynamical ejections, where the more massive BHs have a larger dynamical interaction cross-section \citep{2015ApJ...800....9M,ag2019}. This approach implicitly assumes a 100\% BH retention after supernovae, which is justified given the estimates the of initial mass and escape velocity at formation of 47 Tuc \citep[see][and Section \ref{ssec:bhs}]{2011MNRAS.410.2698G}. For the neutron stars, we make the common assumption of a retention fraction of $10\%$ \citep[e.g.][]{2002ApJ...573..283P}, but our results are actually insensitive to the precise retention fraction of neutron stars adopted (see Section~\ref{results_section}). Our mass function evolution algorithm assigns objects into discrete mass bins, with stars distributed over 30 logarithmically spaced mass bins at time $t=0$. Depending on the BH retention, the mass function consists of 15-19 mass bins at the present day\footnote{See \citet{2017MNRAS.470.2736P} for a discussion of the choice and minimum number of mass bins to ensure fast but stable solutions of multimass {\sc limepy} models.}.
 
Combined with the total cluster mass $M$, the four additional free parameters (power-law exponents $\alpha_1$, $\alpha_2$, $\alpha_{3}$, and $BH_{\rm ret}$) fully define the mass function, allowing to specify all values of $m_j$ and $M_j$ needed to build the multimass model. 

When fitting multimass models similar to the ones described here to mock data from an $N$-body simulation, \citet{2019MNRAS.483.1400H} found that the fraction of the cluster mass in dark remnants and even the mass function of the remnants can be reliably recovered and constrained. Thanks to the trend towards equipartition between objects of different masses, the phase-space distribution of tracer stars is sensitive to the dark remnants, allowing to probe the BH content and the IMF above the present-day main-sequence turnoff mass.

\subsection{Fitting procedure}

For the models described in the previous subsections, we have a total of 12 free parameters, including the 10 following: $W_0$, $g$, $\log{r_{\rm a}}$, $M$, $r_{\rm h}$, $\alpha_1$, $\alpha_3$, $\alpha_3$, $BH_{\rm ret}$, and $\delta$ defined as above. The two additional parameters are nuisance parameters that capture otherwise unaccounted-for observational uncertainties or limitations of the model for the density profile ($s^2$) and the mass function ($F$), as described below.

To compare a given model to the datasets described in Sections \ref{ND} to \ref{MF}, we compute the likelihood of each observational data point given the model parameters. We assume a Gaussian likelihood for all the observables, with the standard devitiation $\delta$ given by the uncertainties (observational, plus in some cases an additional component captured by a nuisance parameter). The likelihood of an observed line-of-sight velocity dispersion $\sigma_{{\rm LOS}*, i}$ with its uncertainty $\delta \sigma_{{\rm LOS}*, i}$ at projected distance $R$ from the centre, given model parameters $\Theta$, is

\begin{equation}
    \begin{array}{l}{\mathcal{L}_{\sigma_{{\rm LOS}, i}}\left(\sigma_{{\rm LOS}*, i}(R), \delta \sigma_{{\rm LOS}*, i}(R) \ \vert \ \Theta \right)=} \\ \displaystyle{\frac{1}{\sqrt{2 \pi}\delta \sigma_{{\rm LOS}*, i}(R)} \exp \left(-\frac{1}{2} \frac{\left(\sigma_{{\rm LOS}, i}(R)-\sigma_{{\rm LOS}*, i}(R)\right)^{2}}{\left(\delta \sigma_{{\rm LOS}*, i}(R)\right)^{2}}\right),}\end{array}
\end{equation}
\noindent{where $\sigma_{{\rm LOS},i}(R)$ is the model line-of-sight velocity dispersion at the distance $R$. Since the line-of-sight velocity dispersion data is dominated by giants, we use the mass bin corresponding to the most massive upper main-sequence stars for the model line-of-sight velocity dispersion to compare with the data\footnote{ For the MUSE data from \citet{2018MNRAS.473.5591K} which are included in the line-of-sight velocity dispersion profile compiled by \citet{2018MNRAS.478.1520B}, this assumption is not entirely accurate. Because of the different stellar densities, longer exposure times were used when observing the outer MUSE fields. Hence, in the outermost bins of the MUSE observations the kinematics trace different stars than in the innermost bins. For 47 Tuc, the difference in the effective stellar mass traced over the spatial extent of the MUSE data is however only $\sim0.1\,\msun$, so this does not have a significant impact on the results.}.}

Similarly, for the {\it HST} proper motion dispersion data in the central regions of the cluster from \citet{2015ApJ...812..149W}, the likelihood of an observed proper motion dispersion in the radial direction $\sigma_{{\rm pmR}*,i}$ with its uncertainty $\delta \sigma_{{\rm pmR}*,i}$ at projected distance $R$ from the centre, given model parameters $\Theta$, is

\begin{equation}
    \begin{array}{l}{\mathcal{L}_{\sigma_{{\rm pmR},i}}\left(\sigma_{{\rm pmR}*,i}(R), \delta \sigma_{{\rm pmR}*,i}(R) \ \vert \ \Theta \right)=} \\ 
    \displaystyle{\frac{1}{\sqrt{2 \pi} \delta \sigma_{{\rm pmR}*,i}(R)} \exp \left(-\frac{1}{2} \frac{\left(\sigma_{{\rm pmR},i}(R)-\sigma_{{\rm pmR}*,i}(R)\right)^{2}}{\left(\delta \sigma_{{\rm pmR}*, i}(R)\right)^{2}}\right),}\end{array}
    \label{L_pmR}
\end{equation}

\noindent{where $\sigma_{{\rm pmR},i}(R)$ is the model proper motion dispersion at the distance $R$ in the radial direction. The likelihood of an observed proper motion dispersion in the tangential direction $\sigma_{{\rm pmT}*, i}$ with its uncertainty $\delta \sigma_{{\rm pmT}*, i}$ at projected distance $R$ from the centre, given model parameters $\Theta$, is}

\begin{equation}
    \begin{array}{l}{\mathcal{L}_{\sigma_{{\rm pmT}, i}}\left(\sigma_{{\rm pmT}*, i}(R), \delta \sigma_{{\rm pmR}*, i}(R)  \ \vert \ \Theta \right) =} \\ 
    \displaystyle{\frac{1}{\sqrt{2 \pi} \delta \sigma_{{\rm pmT}*, i}(R)} \exp \left(-\frac{1}{2} \frac{\left(\sigma_{{\rm pmT}, i}(R)-\sigma_{{\rm pmT}*, i}(R)\right)^{2}}{\left(\delta \sigma_{{\rm pmT}*, i}(R)\right)^{2}}\right),}\end{array}
    \label{L_pmT}
\end{equation}
where $\sigma_{{\rm pmT},i}(R)$ is the model proper motion dispersion at the distance $R$ in the tangential direction. Since these proper motion data only include stars brighter than the main-sequence turn-off, we again use the mass bin corresponding to the most massive upper main-sequence stars for the model proper motion dispersion to compare with the data. Evolved phases like red giant branch (RGB) and  asymptotic giant branch (AGB) stars are expected to have a very similar mass and to behave dynamically like stars with a mass comparable to the mass of turn-off stars. In particular, for 47 Tuc, \citet{2016ApJ...826...88P} conclude that the post-main-sequence mass loss occurs at the end of the AGB phase, so the mass remains constant for stars going through the evolutionary stages from the upper main-sequence up to the horizontal branch. Their slightly higher mass estimates for AGB stars are consistent with the AGB having evolved from somewhat more massive stars.

For the proper motion dispersion observations from \citet{2017ApJ...850..186H}, the likelihood is defined in the same way as in equations (\ref{L_pmR}) and (\ref{L_pmT}). The median mass of the stars in this dataset is 0.38 $\msun$, so we use the mass bin with the corresponding lower mean mass $m_j$ for the model proper motion dispersion to compare with these data, and we refer to the likelihood of an observed proper motion dispersion as $\mathcal{L}_{\sigma_{{\rm pmT \ low},i}}$ and $\mathcal{L}_{\sigma_{{\rm pmR \ low},i}}$, respectively for the radial and tangential components. We note that the median mass of 0.38 $\msun$ is based on a sample of stars from a relatively wide magnitude or mass range ($24>{\rm F814W}>18$; $0.2\gtrsim m/\msun \gtrsim 0.8$), and the velocity dispersion from this whole sample is not necessarily exactly the same as the dispersion of stars with this median mass because the velocities are mass-dependent. We thus also considered a separate test case where we fitted these proper motion dispersion data by using as a reference a lower-mass bin in our models ($m_j = 0.3 \ \msun$), which may be representative of the mean kinematics of the sample considered \citep[cf. Fig. 6 and Fig. 7 from][]{2017ApJ...850..186H}. We conclude in Section \ref{results_section} that this does not significantly affect our results, since the difference in the velocity dispersion is very small ($\lesssim 0.3$~\kms) for this small difference in tracer mass in this outer field.

The number density profile from \citet{2019MNRAS.tmp..651D} is compared to the model prediction for the 
heaviest main-sequence stars. We include in our model density profile a constant background level of 0.08~arcmin$^{-2}$ to match the data in the very outskirts of the cluster. To let the total mass be constrained by the kinematics and stellar mass functions, we only fit on the shape of the profile and not on the absolute values. The likelihood of a measured number density $\Sigma_{*,i}\left(R\right)$ and its uncertainty $\delta\Sigma_{*,i}\left(R\right)$ at the projected distance $R$, given model parameters $\Theta$, is

\begin{equation}
\begin{array}{l} \mathcal{L}_{{\rm ND},i}\left( \Sigma_{*,i}\left(R\right), \delta\Sigma_{*,i}\left(R\right)  \ \vert \ \Theta \right) =  \\
 \displaystyle{\frac{1}{\sqrt{2\pi}\delta\Sigma_{*,i}\left(R\right)} \exp \left(-\frac{1}{2}\frac{\left(K\Sigma_i\left(R\right) -\Sigma_{*,i}\left(R\right)\right)^{2}} {\delta\Sigma_{*,i}\left(R\right)^{2}}\right),}
\end{array}
\end{equation}

\noindent{where $\Sigma_i\left(R\right)$ is the predicted number density from the model at the distance $R$ in the heaviest main-sequence mass bin. $K$ is the scaling parameter that minimizes the vertical offset between the model and observed profile. It re-scales the model profile and allows to fit only on the shape of the number density profile, not its absolute values. It is derived from setting the derivative of the $\chi^2$ expression $\sum_{i=1}^{N_{\rm p}} \left(K \ \Sigma_i\left(R\right) - \Sigma_{*,i}\left(R\right) \right)^2 / \delta\Sigma_{*,i}\left(R\right)^{2}$ with respect to $K$ equal to zero and is defined as
\begin{equation}
K=\frac{\sum_{i=1}^{N_{\rm p}}\Sigma_{*,i}(R)\Sigma_i\left(R\right) /\delta\Sigma_{*,i}\left(R\right)^{2} }{\sum_{i=1}^{N_{\rm p}} (\Sigma_i\left(R\right))^2 / \delta\Sigma_{*,i}\left(R\right)^{2}}
,
\end{equation}

where the sums are over all $N_{\rm {\rm p}}$ data points in the binned observed density profile. In the uncertainty on the observed number density profile ($\delta\Sigma_{*,i}\left(R\right)$), we include both the formal uncertainty from \citet{2019MNRAS.tmp..651D}, $\delta\Sigma_{0,i}\left(R\right)$, and an additional unknown uncertainty encapsulated by nuisance parameter $s^2$ such that $\delta\Sigma_{*,i}\left(R\right)^2 = \delta\Sigma_{0,i}\left(R\right)^2 + s^2$. This nuisance parameter adds a constant component to the uncertainty over the entire extent of the observed number density profile. This effectively allows small deviations between the model and observations in the outer parts of the cluster, for example to account for the effect of potential escapers \citep{2017MNRAS.466.3937C, 2019MNRAS.487..147C} which are not captured by multimass {\sc limepy} models in their current implementation \citep[cf.][]{2019MNRAS.483.1400H}.}

For the comparison of the models and mass function data, we write down the likelihood of an observed number of stars per unit mass $N_{*, i}$ with its uncertainty $\delta N_{*,i}$ at mass $m$ in a given field, given model parameters $\Theta$, as 

\begin{equation}
\begin{array}{l} \mathcal{L}_{{\rm MF},i}\left(N_{*,i}\left(m\right), \delta N_{*,i}\left(m\right)  \ \vert \ \Theta \right) =  \\
 \displaystyle{\frac{1}{\sqrt{2\pi}\delta N_{*,i}\left(m\right)}} \exp \left(-\frac{1}{2}\frac{\left(N_i\left(m\right) - N_{*,i}\left(m\right)\right)^{2}}{\delta N_{*,i}\left(R\right)^{2}}\right),
 \end{array}
\end{equation}
where $N_i (m)$ is the number of stars per unit mass at mass $m$ predicted by the model in the same region. The likelihood for the observed mass function in a given field is the product of the individual likelihoods of all measurements at different masses in that field ($\mathcal{L}_{\rm MF} = \Pi_i \mathcal{L}_{{\rm MF},i}$). We compute the likelihood for the observed mass function given the model parameters in the four different annular regions mentioned above ($\mathcal{L}_{{\rm MF}, 0-0.4^\prime}$, $\mathcal{L}_{{\rm MF}, 0.4-0.8^\prime}$, $\mathcal{L}_{{\rm MF}, 0.8-1.2^\prime}$, $\mathcal{L}_{{\rm MF}, 1.2-1.6^\prime}$). 

In the uncertainty on the observed number of stars per unit mass ($\delta N_{*,i}$), we include both the Poisson error on the number of stars ($\delta N_{0,i}$) and an additional unknown uncertainty captured by the nuisance parameter $F$ and expressed as a fraction between 0 and 1, such that
$\delta N_{*,i}^2 = \delta N_{0,i}^2 + \left(F \ N_{*,i}\right)^2$.  This nuisance parameter encapsulates possible additional sources of uncertainty like errors in the completeness correction of the photometric dataset from which the mass functions were extracted, or limitations of the assumed broken power-law functional form for the underlying global mass function, which may not be a perfect representation of the mass function of the cluster.

For each dataset, we multiply the individual likelihoods for different data points to obtain the likelihood of this dataset given the model. We then multiply the likelihoods for different datasets to obtain the total likelihood:

\begin{equation}
\begin{array}{l} \mathcal{L}_{\rm tot} = \mathcal{L}_{{\rm MF}, 0-0.4^\prime} \ \mathcal{L}_{{\rm MF}, 0.4-0.8^\prime} \ \mathcal{L}_{{\rm MF}, 0.8-1.2^\prime} \ \mathcal{L}_{{\rm MF}, 1.2-1.6^\prime} \\ \mathcal{L}_{{\rm ND}} \ \mathcal{L}_{\sigma_{{\rm pmT}}} \ \mathcal{L}_{\sigma_{{\rm pmR}}} \ \mathcal{L}_{\sigma_{{\rm pmT, low}}} \ \mathcal{L}_{\sigma_{{\rm pmR, low}}} \ \mathcal{L}_{\sigma_{{\rm LOS}}}.
 \end{array}
\end{equation}

Using the total likelihood function above (in practice we maximize the log-likelihood and can conveniently sum individual log-likelihoods for the different datasets to get the total log-likelihood), we determine the posterior distributions of the model parameters using the Markov Chain Monte Carlo sampler {\sc emcee} \citep{2013PASP..125..306F}, adopting uniform priors on all parameters (for the anisotropy radius we actually fit on $\log{(r_{\rm a})}$). The ranges considered for these uniform priors are listed in Table~\ref{best-fit_params}.

\section{Results}
\label{results_section}

\begin{figure*}
\centering
\includegraphics[width=\linewidth]{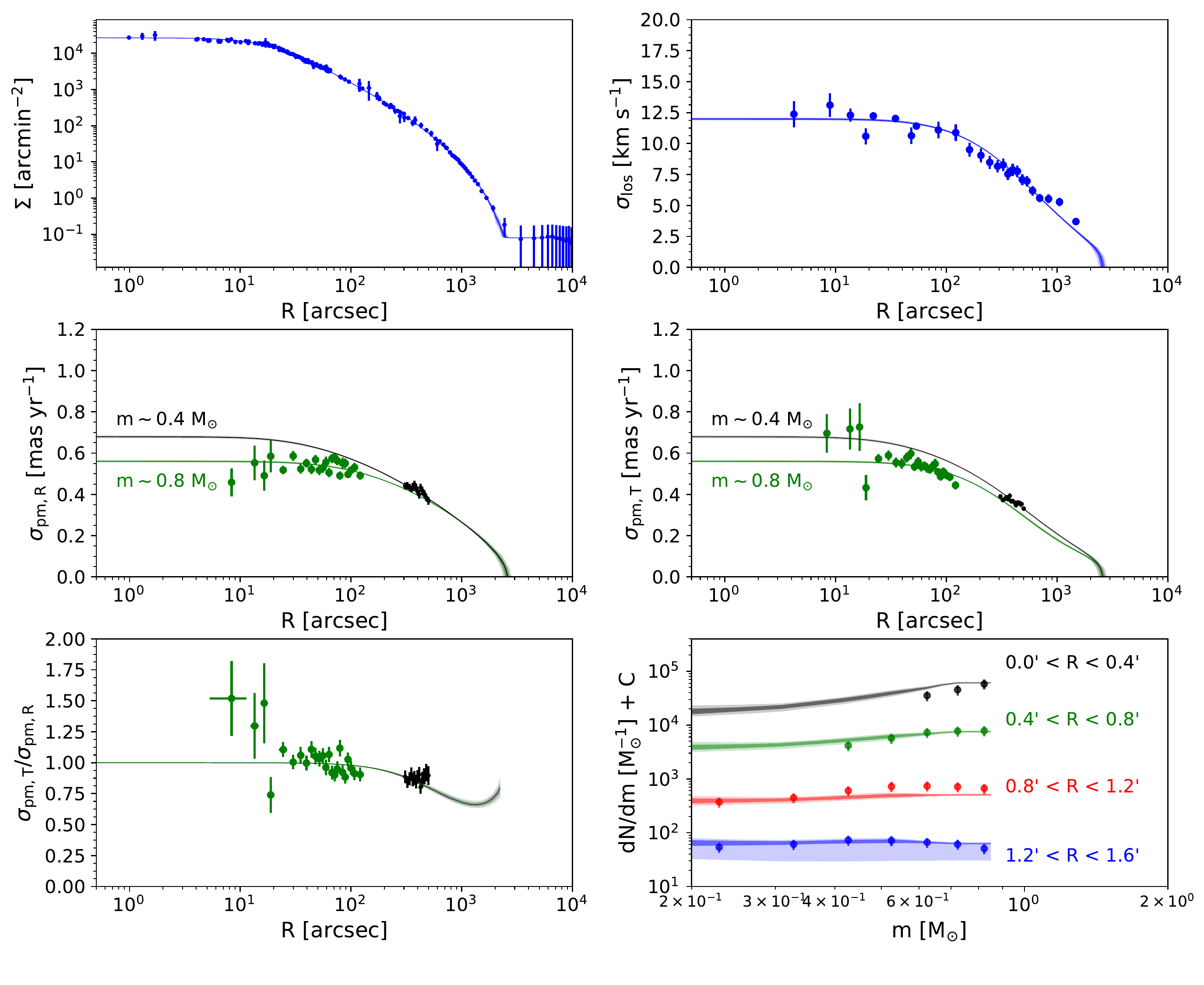}
\caption{Results of our multimass model fit to observations of 47 Tuc. The different datasets are shown with filled circles and error bars, and the best-fitting models with continuous lines. The $1\sigma$ and $2\sigma$ credible intervals of the fitted models are shown with dark and light shaded regions, respectively. {\it Top left:} Number density profile from \citet[][see also Section~\ref{ND}]{2019MNRAS.tmp..651D}. {\it Top right:} Line-of-sight velocity dispersion profile from \citet[][see also Section~\ref{sigma-LOS}]{2018MNRAS.478.1520B}. {\it Middle panels:} proper motion dispersion profiles (radial and tangential components), with innermost data from \citet{2015ApJ...812..149W} in green and outer data from \citet{2017ApJ...850..186H} in black (Section~\ref{PMs}). {\it Bottom left:} ratio of the tangential and radial proper motion dispersions for the datasets shown in the middle panels. {\it Bottom right:} local stellar mass functions at different distances from the cluster centre as described in Section~\ref{MF}. An arbitrary vertical offset has been applied to the the different mass functions in different annuli to facilitate visual comparison of the results.}
\label{fit_profiles}
\end{figure*}

\begin{figure*}
\centering
\includegraphics[width=\linewidth]{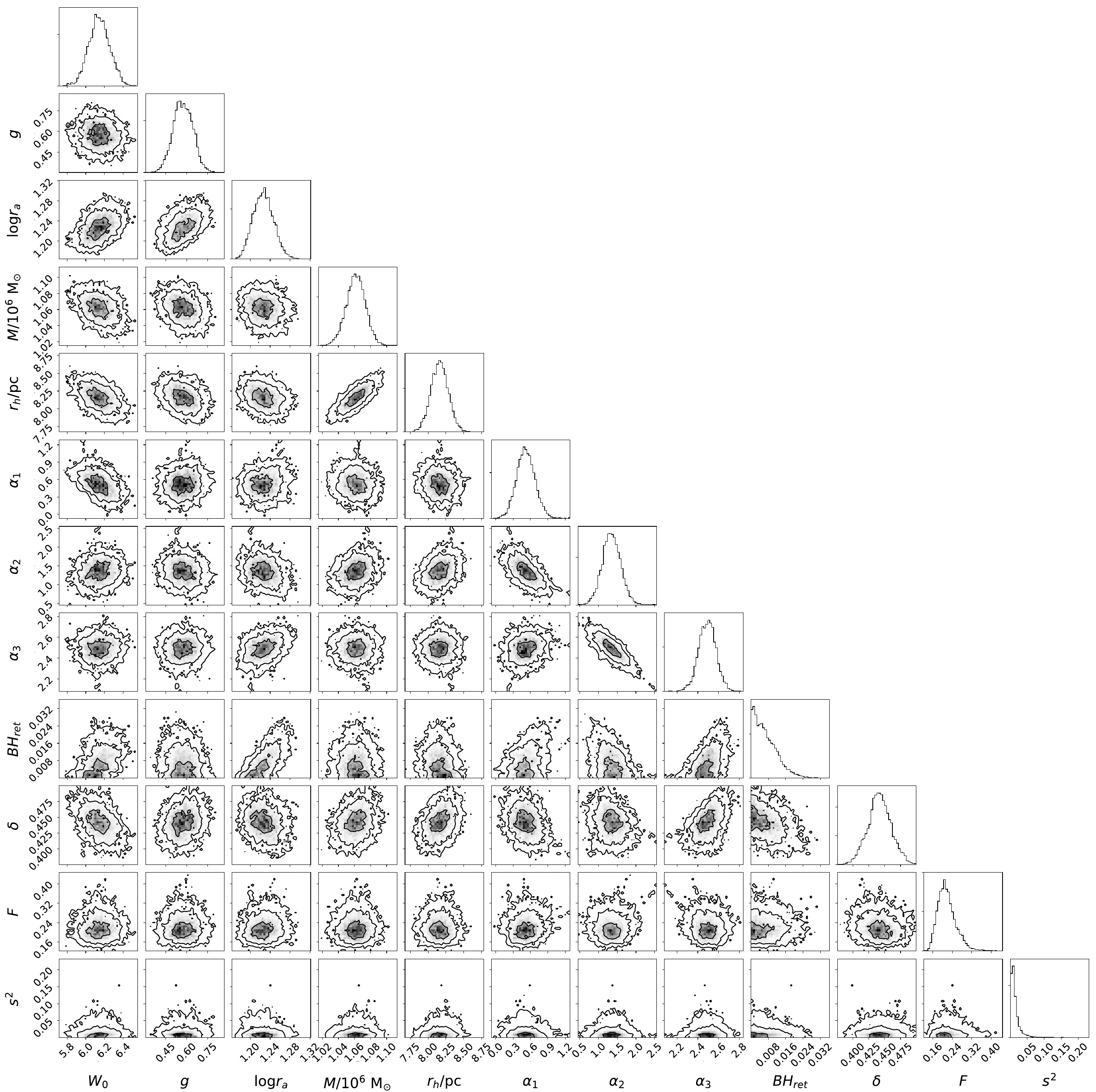}
\caption{Marginalised posterior probability distribution and 2D projections of the posterior probability distribution of the fitting parameters for the {\sc limepy} multimass model fit to 47 Tuc observations.  Contours indicate 1, 2 and 3$\sigma$ levels on the 2D posterior probability distributions.}
\label{triangle}
\end{figure*}

\subsection{Best-fitting model and parameters}

We show in Fig.~\ref{fit_profiles} the resulting model fit to the number density profile, kinematics, and local stellar mass functions. The error bars displayed for the number density profile and mass function data include the error contributions captured by the associated nuisance parameters. Fig.~\ref{triangle} shows the corresponding marginalised posterior probability distribution and 2D projection of the posterior probability distribution for each fitting parameter and pair of parameters. The posterior probability distributions are all well confined within the adopted prior ranges and appear mostly unimodal. The best-fitting model parameters and associated uncertainties (i.e. the median and $\pm 1\sigma$ uncertainties - corresponding to the 16th and 84th percentiles of the marginalised posterior probability distribution) are listed in Table~\ref{best-fit_params}. Note that the best-fitting model (continuous lines) and credible intervals (shaded regions) shown in Fig.~\ref{fit_profiles} do not correspond directly to models computed from the best-fitting values of Table~\ref{best-fit_params}, but rather they are computed from the 16th, 50th, and 84th percentiles (at a given radius or mass) of the profiles computed by sampling parameters from the posterior distribution.

All the different observables considered in the fit are satisfyingly reproduced by the best-fitting model. Some small deviations are found between the best-fitting model and data, but the vast majority are within $2\sigma$, apart from the outermost line-of-sight velocity dispersion data points where the model underestimates the observed velocity dispersion (this is in the regime where we may expect potential escapers, not included in our model, to slightly inflate the velocity dispersion). We note that additional physical ingredients like rotation could eventually be included in the modelling to obtain an even better and complete description of the cluster; 47 Tuc indeed displays significant rotation \citep[e.g.][]{2017ApJ...844..167B}. Such improvements are however beyond the scope of this study.

All our fitting parameters are well constrained by the data, but Fig.~\ref{triangle} reveals small degeneracies between some pairs of parameters (see oval shapes in the 2D projections of the posterior probability distributions). For example, $\alpha_1$ and $\alpha_2$ show an anti-correlation in the sense that a shallower low-mass slope ($\alpha_1$) can be compensated by steeper intermediate slope ($\alpha_2$). A similar behaviour is seen for $\alpha_2$ and $\alpha_3$. This perhaps suggests that $\alpha_2$ is somewhat redundant and that we could have approximated the mass function of visible stars with a single power law, but given that the three mass function slopes are well constrained we stick to the adopted functional form. $\alpha_3$ and $BH_{\rm ret}$ are also degenerate: a flatter high-mass mass function slope produces more BHs from the IMF, so this has to be accompanied by a lower $BH_{\rm ret}$ to yield the same number of BHs (or mass in BHs) after ejection. $\alpha_1$ is similarly degenerate with $BH_{\rm ret}$, which is more difficult to interpret. Possibly it is because, as the low-mass end of the mass function slope gets steeper, a larger $BH_{\rm ret}$ is needed to produce the same number of BHs per unit cluster mass. Also, for flatter MFs at low masses, the more massive stars and remnants do not segregate as far into the centre as for steeper slopes \citep{2015MNRAS.448L..94S}, hence more BHs are needed to raise the central mass density and velocity dispersion by a certain amount. $BH_{\rm ret}$ is also degenerate with the amount of anisotropy, a consequence of the well-known mass-anisotropy degeneracy. Both a larger retention of black holes and a larger amount of radial anisotropy  increases the central velocity dispersion \citep[e.g.][]{2017MNRAS.468.4429Z, 2019MNRAS.482.4713Z}.

Our best-fitting mass segregation parameter ($\delta = 0.44\pm0.02$) is close to the commonly adopted value of $\delta=0.5$. The resulting relation between the central velocity dispersion and mass (in the range $\sim0.4-0.8 \ \msun$) is shallower ($\sigma(r=0)\propto m^{-0.38}$). In projection our model gives: $\sigma_{\rm p}(R=0) \propto m^{-0.32}$.

{ Comparing to the results of \citet{2018MNRAS.478.1520B} who fitted their grid of $N$-body models to similar data for 47 Tuc, we find a half-mass radius that is $\sim30\%$ larger and a total cluster mass that is $\sim25\%$ larger (both significantly different than allowed by the statistical errors on the fits of the two studies). We note that the ratio $M/r_{\rm h}$, which is proportional to the velocity dispersion squared, is very similar in both models. Part of the difference in mass may be traced to the shallower present-day mass function inferred by these authors, who find a power-law slope of $\alpha=-0.53$ over the mass range $0.2 \ \msun < m < 0.8 \ \msun$. Our mass function, with a break at  $0.5 \ \msun$, is comparable below $0.5 \ \msun$ but steeper above this break mass, and thus contains a larger proportion of low-mass stars which would be preferentially distributed in the outer parts of the cluster and could increase the inferred total mass and half-mass radius without significantly affecting the mass profile in the inner regions. Moreover, our total likelihood function includes a comparison with radial and tangential proper-motion data that provide constraints on the radial anisotropy in 47 Tuc, whereas such data is not included in the fits of \citet{2018MNRAS.478.1520B} - anisotropy cannot be freely varied in their grid of $N$-body models. It is possible that their best-fitting model has too much radial anisotropy, which would then require less mass to fit the central velocity dispersion and could lead to underestimating the cluster mass or their models have less mass in the outer parts. An other explanation could be that the visible stars are more segregated towards the centre in our models because we allowed $\delta$ to be free. A direct comparison between the models would be required to understand the differences.}

\begin{table}
\caption{Best-fitting model parameters and associated uncertainties (i.e. the median and $\pm 1 \ \sigma$ uncertainties - corresponding to the 16th and 84th percentiles of the marginalised posterior probability distribution) for our multimass model fit to 47 Tuc data.}
\centering
\label{best-fit_params}
\renewcommand{\arraystretch}{1.6}\addtolength{\tabcolsep}{4.6pt}
\begin{tabular}{ccc}
\hline
Parameter        & Value & Prior range \\
\hline
$W_0$	         & $6.1^{+0.1}_{-0.1}$ & [3, 20] \\
$g$		         & $0.57^{+0.07}_{-0.07}$ & [0, 2.3] \\
$\log{r_{\rm a}/\rm{pc}}$		         & $1.23^{+0.02}_{-0.02}$ & [0, 5] \\
$M/10^6 \ \msun$		         & $1.06^{+0.01}_{-0.01}$ & [0.01, 10]  \\
$r_{\rm h}/\rm{pc}$	         & $8.16^{+0.12}_{-0.12}$ & [0.5, 15] \\

$\alpha_{1}$	         & $0.52^{+0.17}_{-0.16}$ & [-2, 6] \\
$\alpha_{2}$	         & $1.35^{+0.25}_{-0.23}$ & [-2, 6] \\
$\alpha_{3}$	         & $2.49^{+0.08}_{-0.08}$ & [-2, 6] \\
$BH_{\rm ret}$	         & $0.59^{+0.06}_{-0.04} \ \%$ & [0, 100\%] \\
$\delta$	         & $0.44^{+0.02}_{-0.02}$ & [0.3, 0.5] \\
$F$	         & $0.21^{+0.04}_{-0.03}$ & [0, 0.5] \\
$s^2$	         & $0.01^{+0.01}_{-0.005}$ & [0, 10] \\
\hline
\end{tabular}
\end{table}

\begin{figure}
\begin{center}
\includegraphics[width=\columnwidth]{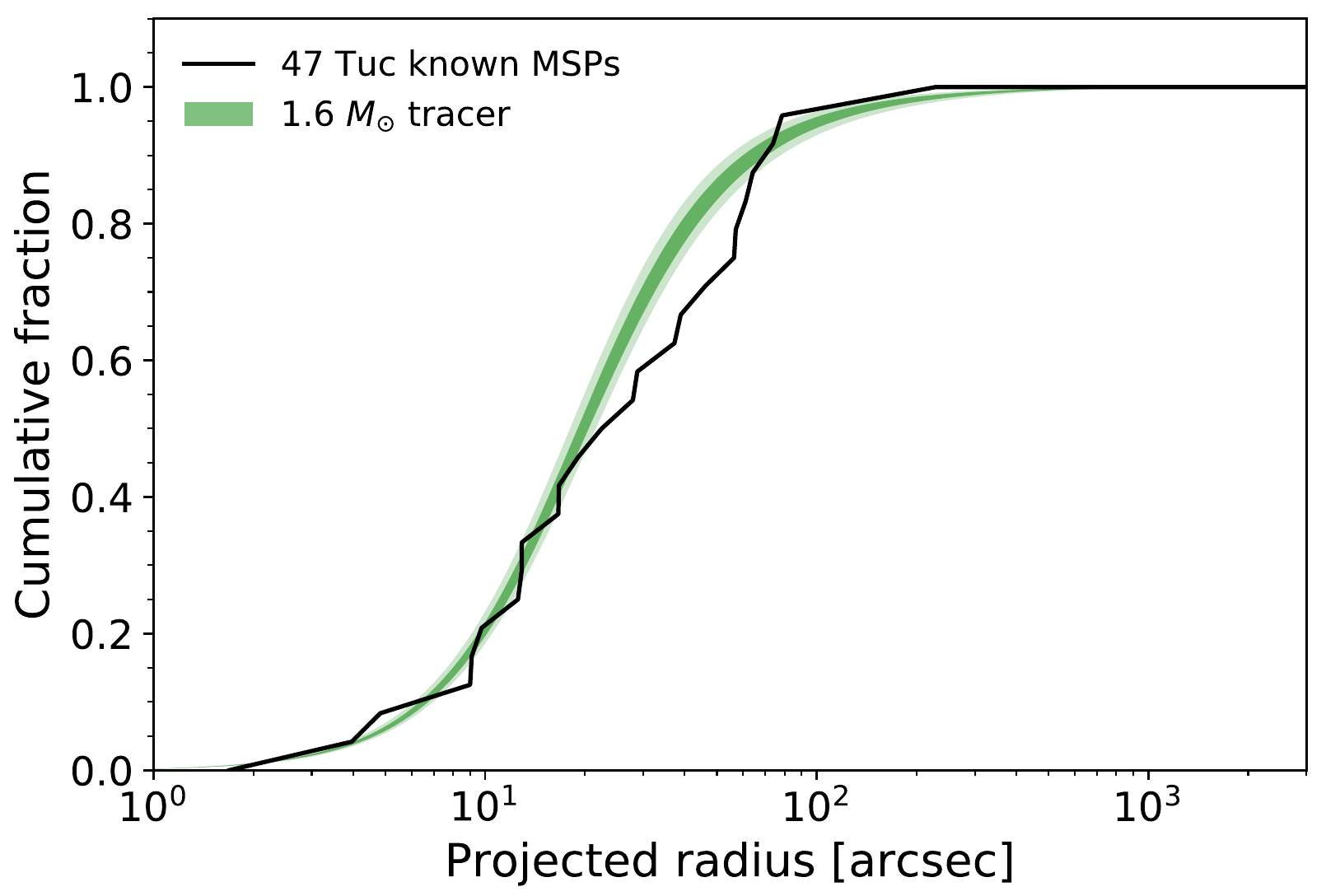}
\caption{Cumulative radial distribution of the 25 known MSPs in 47 Tuc (black line) compared to the prediction of our best-fitting multimass models for a 1.6 $\msun$ tracer (the typical mass of MSP systems in 47 Tuc given the presence of binary companions in a fraction of the systems). The $1\sigma$ and $2\sigma$ credible intervals of the model distribution for these objects are shown with dark and light shaded green regions, respectively.}
\label{MSP_radial}
\end{center}
\end{figure}

\subsection{Radial distribution and line-of-sight accelerations of millisecond pulsars}

Fig.~\ref{MSP_radial} shows the cumulative radial distribution of the 25 known MSPs \citep{2017MNRAS.471..857F, 2016MNRAS.462.2918R, 2018MNRAS.476.4794F} in 47 Tuc (black line) compared to the prediction of our best-fitting multimass models for a 1.6 $\msun$ tracer (the typical mass of MSP systems in 47 Tuc given the presence of binary companions in a fraction of the systems. We assume a mass of 1.4 $\msun$ for the isolated pulsars). We emphasize that these data were not included in our model fit and that only observables probing the phase-space distribution of visible stars below $\sim 0.85 \ \msun$ were used to constrain the free parameters of the model. Thus, the remarkable agreement between the data and model predictions lends further support to the ability of our models to reliably reproduce the underlying distribution of objects of different masses, including heavy dark remnants.

As discussed in Section~\ref{Section_intro}, long-term timing of millisecond pulsars can also be used to probe the gravitational potential of a globular cluster. Ignoring intrinsic effects modifying the period (spin or orbital) of the MSP, this period and its derivative can be linked to the line-of-sight gravitational acceleration within the cluster potential as $\dot{P}/P = a_{\rm los}/c$ \citep{1993ASPC...50..141P, 2017ApJ...845..148P}, where $c$ is the speed of light (a positive $a_{\rm los}$, i.e. away from the observer, leads to an apparent spin-down or growth of the period). In the presence of an unknown intrinsic spin-down (e.g. due to magnetic breaking in a millisecond pulsar), the inferred acceleration is an upper limit to the true acceleration from the cluster potential. If the intrinsic component is known, the acceleration can be computed as $a_{\rm los}/c =  \left(\dot{P}/P\right)_{\rm meas} - \left(\dot{P}/P\right)_{\rm int}$, where the `meas' and `int' subscripts refer to the measured and intrinsic period derivatives. Additional terms, for example the acceleration due to the Galactic potential and the proper motion, are usually negligible \citep{1993ASPC...50..141P}.

\begin{figure}
\centering
\includegraphics[width=\columnwidth]{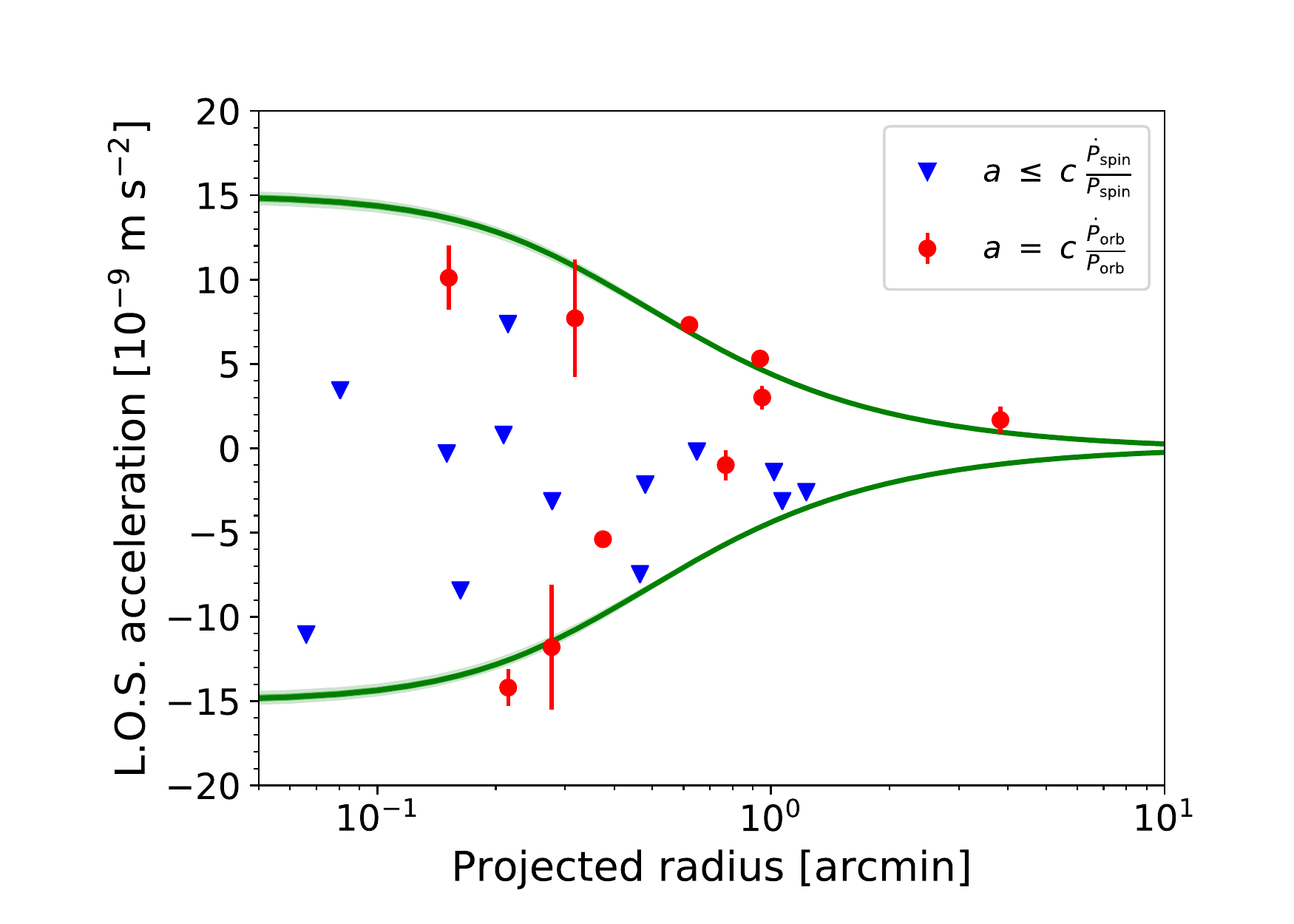}
\caption{Line-of-sight gravitational acceleration of pulsars as a function of projected distance from the centre of 47~Tuc. The green envelope bounds the maximum (positive) and minimum (negative) line-of-sight acceleration at a given projected distance from the centre, based on the mass distribution of our best-fitting multimass model. Red circles show line-of-sight accelerations inferred from the measured orbital period derivative of 10 MSPs, while blue inverted triangles show upper limits on the line-of-sight acceleration in the field of the cluster for another 13 MSPs based on their spin-period derivative (which could have an intrinsic spin-down component, hence the upper limit on the real acceleration in the field of the cluster).}
\label{acc_pulsars}
\end{figure}

Fig.~\ref{acc_pulsars} compares inferred line-of-sight accelerations for MSPs in 47 Tuc to the maximum (positive) and minimum (negative) line-of-sight accelerations predicted by our best-fitting multimass model as a function of projected distance from the centre of the cluster. We split the sample of inferred accelerations into two subgroups following the data presented in Section~\ref{pulsar_data}: (1) line-of-sight accelerations determined from the orbital period derivatives of 10 MSPs in binary systems, which we assume are directly probing the acceleration in the cluster because the intrinsic orbital period derivative should be negligible, and (2) upper limits on the line-of-sight accelerations based on the spin period derivatives of the other 13 MSPs, which could have a non-negligible (but currently unknown) intrinsic spin-down component, hence the upper limit. We see from Fig.~\ref{acc_pulsars} that all upper limits from the latter subgroup are safely above the minimum (negative) line-of-sight acceleration allowed by our best-fitting model and scattered between the maximum (positive) and minimum (negative) acceleration boundaries. The true accelerations determined for the 10 binary systems are also in good agreement with our model. They all fall within the boundaries allowed by the model within 1$\sigma$, except for one system (47 Tuc S) which is however still consistent within less than 1.5$\sigma$. The majority of the accelerations for these binary systems cluster around the maximum and minimum boundaries. We note that this is not unexpected; the probability distribution of the acceleration at a given projected distance from the centre peaks near these boundaries as a result of the density distribution of the cluster and the variation with line-of-sight distance of the component of the acceleration vector that is projected along the line of sight. 
The good agreement between the MSP data and our mass model lends confidence to our choice of model and associated assumptions.
A more in-depth statistical analysis of the pulsar accelerations and associated constraints on the mass profile and BH content of 47 Tuc will be presented in a future work.

\begin{figure*}
\centering
\includegraphics[width=\linewidth]{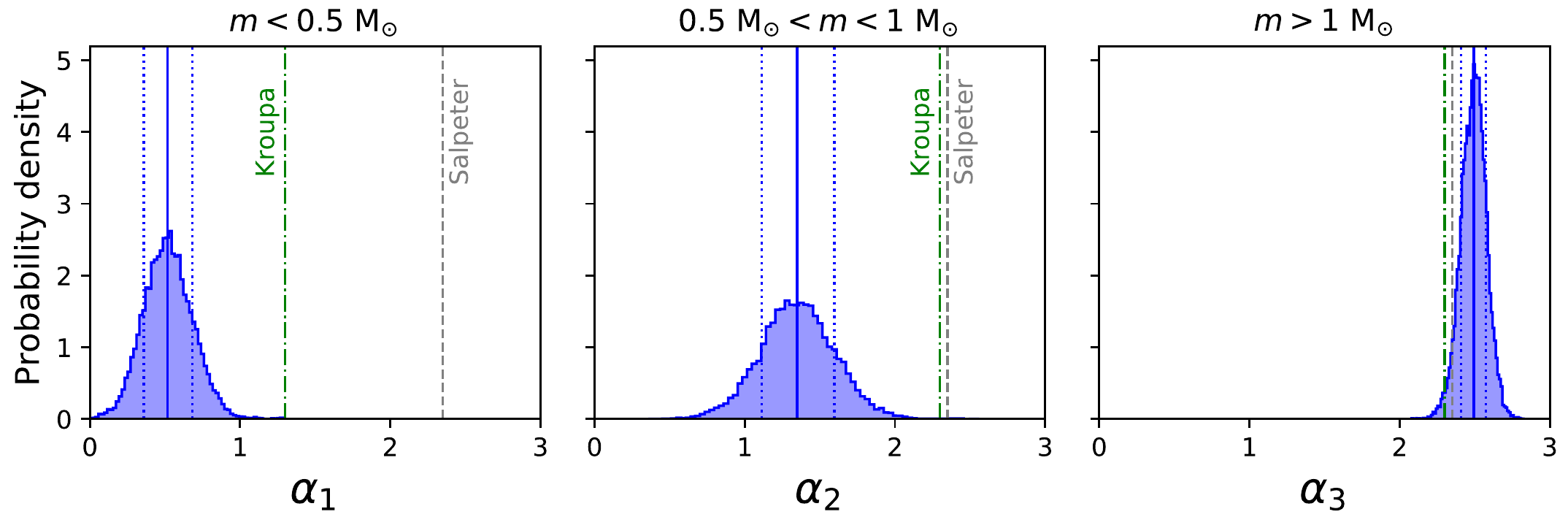}
\caption{Marginalised posterior probability distribution (blue shaded histograms) for each of the three power-law exponents defining the (initial) stellar mass function in 47 Tuc, for the ranges $m<0.5 \ \msun$ ($\alpha_1$), $0.5<m<1 \ \msun$ ($\alpha_2$), and $m>1 \ \msun$ ($\alpha_3$). For each distribution, the median is shown by a solid blue vertical line and the $1\sigma$ uncertainties (16th and 84th percentiles) by the dashed dotted blue lines. For reference, we also show vertical lines with the slopes for the Kroupa (dash-dotted green lines) and Salpeter (grey dashed lines) in the corresponding mass range. The tension with a standard IMF is clear in the low-mass regime.
}
\label{alpha_pdf}
\end{figure*}

\subsection{Global present-day mass function and IMF}

We highlight in Fig.~\ref{alpha_pdf} the marginalized posterior probability distribution for each of the three power-law exponents defining the (initial) stellar mass function in 47 Tuc. The inferred global mass function for low-mass stars is significantly flatter than a canonical IMF \citep{2001MNRAS.322..231K}, with $\alpha_1=0.52^{+0.17}_{-0.16}$ and $\alpha_2=1.35^{+0.25}_{-0.23}$ compared to $\alpha_1=1.3$ and $\alpha_2=2.3$ for the Kroupa IMF in the same mass range. Despite allowing the high-mass initial mass function slope to be completely free to vary, we are nevertheless able to constrain the value of $\alpha_3$ to be $2.49^{+0.08}_{-0.08}$, slightly steeper but consistent with the Salpeter value \citep[2.35;][]{1955ApJ...121..161S} within 1.5$\sigma$. It is worth nothing that the uncertainty on $\alpha_3$ is smaller than that of $\alpha_1$ and $\alpha_2$, i.e. despite the fact that the number of low-mass stars can to a large extent simply be counted, the inferred high-mass slope is better constrained. This is because of the sensitivity of the central velocity dispersion to the central mass distribution, which is dominated by dark stellar remnants. Our method provides a unique way to probe the stellar IMF at high redshift (for a given choice of the initial-final mass relation).

\subsection{Model assumptions and uncertainties}
\label{ssec:assumptions}
In addition to the results presented above, for which a distance of 4.45 kpc was adopted \citep{2018ApJ...867..132C}, we also repeated the fitting procedure by assuming distance values of $D=4.2$~kpc and $D=4.7$~kpc (2$\sigma$ excursions from the value of \citealt{2018ApJ...867..132C}). This mainly affected the inferred mass, radius, and some of the mass function slopes, although not in a major way. All other model parameters and inferred secondary quantities (including the mass in BHs; see discussion below) remained the same within 1$\sigma$ uncertainties. For $D=4.2$~kpc, the inferred $r_{\rm h}$ decreased to $7.4\pm0.1$~pc and the inferred cluster mass decreased to $0.88\pm0.01 \times10^6 \ \msun$. On the other hand, for $D=4.7$~kpc, the inferred $r_{\rm h}$ increased to $8.7\pm0.1$~pc and the inferred cluster mass increased to $1.20\pm0.01 \times10^6 \ \msun$. Either way, the effect on the radius is less than 10\% and the effect on 
the cluster mass less than 20\%. While $\alpha_1$ changed by less than 0.1 when adopting the two extreme distance values, the other power-law exponents were more significantly affected: $\alpha_2=1.74^{+0.27}_{-0.29}$ and $\alpha_3=2.21^{+0.10}_{-0.11}$ for $D=4.2$~kpc, whereas $\alpha_2=1.06^{+0.29}_{-0.25}$ and $\alpha_3=2.64^{+0.07}_{-0.09}$ for $D=4.7$~kpc. Our conclusion that the low-mass mass function is significantly flatter than a Kroupa IMF is therefore robust and not strongly affected by uncertainties on the distance. 

We also performed fits assuming different retention fractions for the neutron stars, with anywhere from zero to several thousands of neutron stars retained, and found that all the best-fitting parameters remained unchanged within 1$\sigma$ uncertainties. Similarly, adopting a lower mass of 0.3~$\msun$ (compared to 0.38~$\msun$) for the model tracer stars to be compared with the proper motion data from \citet{2017ApJ...850..186H} did not significantly change any of the best-fitting parameter values.

We finally recall that we have assumed that modification of the mass function by dynamical evolution and preferential escape of low-mass stars and remnants is negligible for 47 Tuc. When varying the mass function at low masses, we therefore assume that this corresponds to variations in the IMF. Note that if the IMF was actually normal and was dynamically depleted of low-mass stars by some dynamical effect not considered, then the low-mass end of the white dwarf mass function would also be affected \citep[e.g. see Fig. 2 from][]{2018MNRAS.473.4832G}. This may require a somewhat flatter high-mass IMF to end up with the same mass in remnants, but it is not immediately clear precisely how this would affect the mass model.

\section{Discussion}
\label{discussion_section}

\subsection{Black holes in 47 Tuc}
\subsubsection{No need for an IMBH}

Our best-fitting multimass model of 47 Tuc can naturally accommodate the accelerations of MSPs in this cluster, without invoking the presence of a central IMBH. The good agreement between the model and a range of observational constraints suggests that an IMBH is not needed given the data currently available \citep[see also][]{2019ApJ...875....1M}. \citet{2017Natur.542..203K} reached a different conclusion when using the spin-period derivatives of pulsars (along with an assumed intrinsic spin-down distribution to infer accelerations) as the sole observable to distinguish between models with and without an IMBH; their models favoured an IMBH with a mass $M_{\bullet}\sim 2200^{+1500}_{-800} \ \msun$. It is unclear which assumption(s) in their analysis (see Section~\ref{Section_intro}) could have led to this different result, but we note that it is based on a grid of isolated $N$-body models that would have a much steeper present-day mass function, closer to a \citet{2001MNRAS.322..231K} IMF and the assumption of a shorter distance. In Section~\ref{ssec:assumptions} we showed that adopting a shorter distance leads to a shallower IMF slopes at high masses. This suggests that the data then prefers additional dark mass in the centre in the form of remnants, which might explain the need for an IMBH in the analysis of \citet{2017Natur.542..203K}.

\begin{figure}
\centering
\includegraphics[width=\columnwidth]{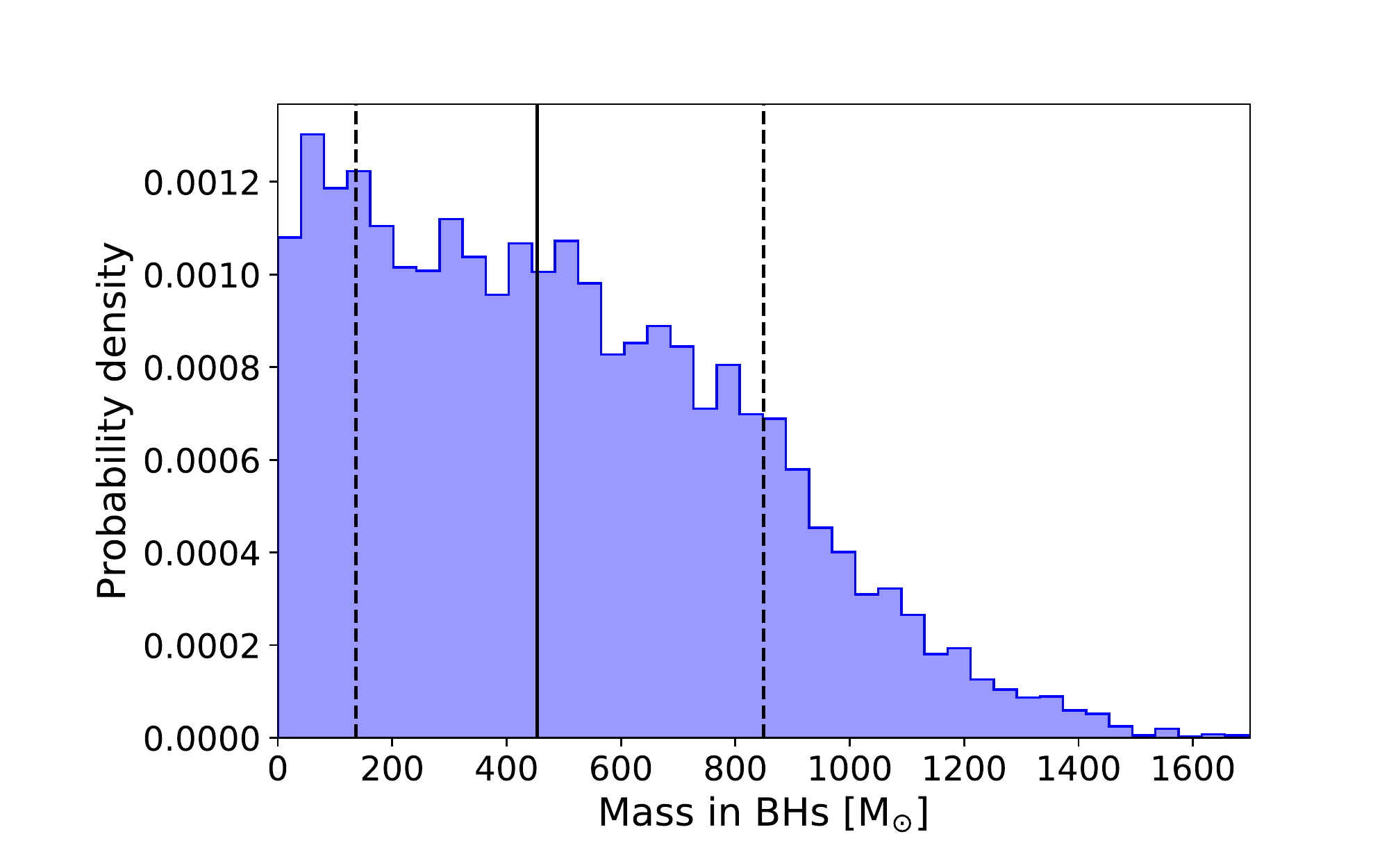}
\caption{Posterior probability distribution for the total mass in BHs in our multimass model of 47 Tuc. The median is shown by the solid black line and the $1\sigma$ uncertainties (16th and 84th percentiles) by the dashed black lines.}
\label{BH_posterior}
\end{figure}

\subsubsection{Constraints on the stellar-mass BH population}
\label{ssec:bhs}

From the constraints on the mass function, total cluster mass and black hole retention fraction provided by our multimass model fit, we can recover the probability distribution for the total mass in BHs at the present day in 47 Tuc (Fig.~\ref{BH_posterior}). We find a preference for a population of BHs with a total mass of $430^{+386}_{-301}$~$\msun$, or equivalently  $141^{+97}_{-95}$ BHs with a mean mass of $3.1\pm0.4 \ \msun$ given our assumptions impacting the BH mass function (where our average BH mass is low because we assume massive BHs are dynamically ejected). The BHs contribute $\sim10^4 \ \msun$~pc$^{-3}$ to the central mass density of the cluster, which we infer to be $\sim10^5 \ \msun$~pc$^{-3}$ from our best-fitting models. While the BH contribution to the central density is not dominant, it is much more important (thanks to mass segregation) than one might guess from the very small fraction of the cluster mass in BHs ($\lesssim0.1\%$). This makes our models sensitive to a relatively small BH population and allows us to place the useful constraints above.

Our results are also consistent with a negligible number of BHs within $\sim1.5\sigma$. This is in agreement with the 19 BHs retained in the Monte Carlo model of 47~Tuc by \citet{2011MNRAS.410.2698G}, although we note that their results are based on a limited number of Monte Carlo models (the study indeed does not claim a full exploration of the parameter space for the cluster initial conditions and the retention of BHs). { We should mention that these Monte Carlo models assumed that BHs and neutron stars received the same (large) natal kicks, naturally leading to a small number of retained BHs early on}. { Our constraints on the BH content in 47 Tuc are also in keeping with the low number of retained BHs implied by the results of \citet{2018MNRAS.478.1844A} and \citet{2018MNRAS.479.4652A} and} in agreement with the population of $\sim20$~BHs (but possibly as large as $\sim150$ BHs within 2$\sigma$) inferred by \citet{2018ApJ...864...13W} based on the observed mass segregation between giants and main-sequence stars in the central regions of the cluster, although we note that our models are simultaneously fitted to several additional observables. To model the velocity dispersion in the core of 47 Tuc with the main goal of placing constraints on the presence of an IMBH, \citet{2019ApJ...875....1M} took into consideration populations of objects more centrally concentrated than the typical main-sequence stars (binaries and stellar remnants) which may cause the velocity dispersion to rise in the core and affect the inferred IMBH mass. They found that concentrated populations of binary stars and dark stellar remnants alone appear to be enough to explain the velocity dispersion in the core, and thus that there is no evidence for an IMBH (unless an unrealistically small population of remnants is assumed). Pushing their adopted retention of neutron stars and BHs higher than $\sim8.5\%$ even led to a larger velocity dispersion than is observed in the core, requiring an unphysical negative IMBH mass in their Jeans model fit. Given their assumptions about the initial BH population and their adopted object mass of $10 \ \msun$ for each BH, this maximum retention fraction translates to an upper limit of $1615 \ \msun$ in BHs, consistent with our new constraints. All these results are in keeping with the presence of a strong BH binary candidate in 47 Tuc \citep[X 9;][]{2015MNRAS.453.3918M, 2017MNRAS.467.2199B, 2017ApJ...851L...4C, 2018MNRAS.476.1889T}, consistent with the idea that 47 Tuc still hosts at least some BHs today.

Forcing our models to retain a significantly larger population of BHs would adversely affect the quality of the fit by producing a core that is too large and suppressing mass segregation to the extent that the observed change in the local stellar mass function slope with radius cannot be matched. As most BHs formed in 47~Tuc are expected to have been ejected dynamically (see below), the inferred upper limit provides important constraints on the dynamical evolution history of the cluster.

Due to our adopted prescription for the initial-final mass relation of massive stars, our assumption that the more massive black holes are ejected first, and the low inferred low dynamical retention fraction of BHs, the present-day population of BHs that we infer in 47 Tuc is comprised of objects of a few solar masses only and at most $\sim6$~M$_{\odot}$ (see Fig.~\ref{PDMF_IMF}). If the retained BHs were in reality more massive on average, our reported number of retained BHs would overestimate the true number of BHs and should then be taken as an upper limit (if the BHs are more massive, fewer are needed to produce the same dynamical effect). This would be the case if a significant mass gap exists between the heaviest neutron stars and the lightest BHs \citep[with no BH below $\sim5 \ \msun$, e.g.][]{2012ApJ...757...91B}, which is not captured in our adopted initial-final mass relation (our mininum BH mass is $2.6 \ \msun$).

Our assumption that the massive BHs are preferentially removed (as expected from dynamical ejections) could possibly also lead to a similar bias. A fraction of the BHs formed in a GC is expected to leave the cluster due to experiencing natal kicks larger than the escape velocity at the time of supernova. This is expected to lead to a mass-dependent retention fraction and predominantly remove low-mass BHs. However, the present-day central escape velocity of our model is $\gtrsim50\,$\kms \, and this would have been at least a factor of two higher at birth because of stellar mass loss and subsequent adiabatic expansion. If we assume that BHs receive the same momentum kick as neutron stars, then \citet{ag2019} showed that for escape velocities above $\sim100$~\kms \ almost all BHs are retained (their Fig.~1). { Note however that according to the mass fallback prescription, low-mass BHs will receive larger natal kicks and it would be easier for them to escape the system. The argument above about the escape velocity might be softened because initial mass loss associated with the most massive stars will lower the central potential before lower-mass BHs form. This would lead to a larger average BH mass at the present day. On the other hand, the initial escape velocity could have been substantially higher than the conservative estimate obtained from correcting for adiabatic expansion only, as we will discuss in Section~\ref{ssec:bhs}.}

{ An exciting prospect to break the degeneracy between dynamically ejected BHs and those lost via natal kicks is by} finding BH candidates with reliable mass estimates, either with multi-epoch spectroscopy if they are in a detached binary with a stellar companion \citep{2018MNRAS.475L..15G}, or microlensing of  background (SMC) stars or quasars \citep{2016MNRAS.458.3012W,2019arXiv190407789W}. This would provide constraints on the BH mass function in our mass models and therefore on natal kicks and dynamical BH ejection{, because a small dynamical retention fraction implies low BH masses, while a small retention fraction resulting from natal kicks leads to a high BH masses (if natal kicks are larger for low-mass BHs). }

\subsubsection{The effect of binaries}

We did not explicitly include binary systems when building the global present-day mass function for our multimass models, and it is worth discussing the potential effect of ignoring these binaries. From an analysis on the visual/near-infrared colour-magnitude diagram (CMD) of 47~Tuc stars, \citet{2019ApJ...875....1M} estimated binary fractions in different mass-ratio ($q$) ranges (above $q=0.5$, where binaries can be isolated from the single-star main sequence). For the ranges, $0.5<q<0.7$, $0.7<q<0.9$, and $q>0.9$, they find binary fractions of 2.65\%, 0.98\%, and 0.3\%, respectively.

Only binaries with a primary mass close to the main-sequence turnoff mass and a $q$ above $\sim0.7$ would have a system mass comparable to neutron stars. From the estimates in Table 2 of \citet{2019ApJ...875....1M}, the binaries with a system mass above $1.4 \ \msun$ could make up to $\sim5000 \ \msun$ of the cluster mass. This is less than 1\% of the total cluster mass but these objects would be concentrated in the inner regions of the cluster due to mass segregation and could provide a significant contribution to the gravitational potential there. However, in our tests where we changed the adopted retention fraction of neutron stars, we found that retaining anywhere from very few to several thousands of neutron stars did not significantly affect our results, in particular with respect to the mass in BHs and IMF. Neutron stars (and heavy binaries with comparable mass) are thus expected to be relatively unimportant dynamically. Their total mass contribution would actually be dwarfed by the mass in massive ($>1 \ \msun$) white dwarfs in our best-fitting model, which add up to $\sim6\times10^4 \ \msun$.

Interestingly, from their CMD analysis, \citet{2019ApJ...875....1M} estimate a total mass of $\sim 41,500 \ \msun$ in heavy binaries ($>1 \ \msun$), and their assumptions about the mass function and retention of stellar remnants lead to a mass of $\sim 27,600 \ \msun$ in heavy remnants (massive white dwarfs $>1 \ \msun$, neutron stars, BHs) in their model. Their total mass in objects more massive than $1 \ \msun$ ($\sim 6.9\times10^4 \ \msun$) is therefore very comparable to what we infer in our multimass model. By ignoring heavy binaries when building our multimass model, there is therefore a possibility that this missing mass is compensated by favouring more mass in heavy remnants and a model with a high-mass IMF ($\alpha_3$) that is shallower than the true cluster IMF. In that respect, it is interesting to recall that the Monte Carlo model of 47~Tuc by \citet{2011MNRAS.410.2698G} required a relatively steep IMF for stars above the main-sequence turnoff, with an index of about 2.8 (compared to 2.35 for a Salpeter mass function), but still a relatively flat IMF for the lower main sequence ($\alpha_1=0.4$).

The vast majority of (undetected) low-$q$ binaries ($q<0.5$) would have total binary masses below $1\ \msun$ and behave dynamically like the general main-sequence population, but they could potentially bias the star counts as a function of mass in the observed local stellar mass functions. However, even under the conservative assumption than only a quarter of binaries have $q>0.5$, the total binary fraction is expected to be $<10\%$ in 47 Tuc \citep[e.g.][]{2012A&A...540A..16M} and we do not expect this small population of low-$q$ binaries to affect our conclusions. We defer a detailed treatment of the effect of binaries on our models to future work.

We note that \citet{2019ApJ...875....1M} inferred a total cluster mass and a central velocity dispersion that are significantly larger than what other studies have found for 47 Tuc, and this may have led them to overestimate the total mass in binaries and in heavy remnants. They obtain a total cluster mass of $1.47\times10^6 \ \msun$, at least $\sim35\%$ higher than what we (Table~\ref{best-fit_params}), \citet{2011MNRAS.410.2698G}, and \citet{2018MNRAS.478.1520B} found. This could partly be due to a limitation of their method, which only fits on kinematic data in the central region of the cluster and does not yield a self-consistent model of different observables (e.g. not including the density distribution) over the whole spatial extent of the cluster. Their central velocity dispersion ($\sim 14-15$~\kms) is also high compared to the value of $\sim 12$~\kms\ that is derived from the LOS velocities and that we find from our fit to the the proper motion and line-of-sight kinematic data of \citet{2018MNRAS.478.1520B} and \citet{2015ApJ...812..149W}. It is not clear what the range of magnitudes (and corresponding masses) of the stars entering the central velocity dispersion profile of \citet{2019ApJ...875....1M} is, but if stars fainter than the main-sequence turnoff (thus lower-mass tracers) were included, it may explain the discrepancy; the kinematic data that we used are based on samples of stars with a mass comparable to turnoff stars. An indication in that sense comes from the central proper motion velocity dispersion measured by \citet{2006ApJS..166..249M} for stars with a mass comparable to the turnoff mass, $0.609\pm0.010$~mas yr$^{-1}$, which corresponds to $12.8$~\kms\ when scaled to a distance of 4.45 kpc, significantly lower than the value of $\sim 14-15$~\kms\ reported by \citet{2019ApJ...875....1M}.

\subsection{The initial stellar mass function}
\label{IMF}

\begin{figure}
\centering
\includegraphics[width=\columnwidth]{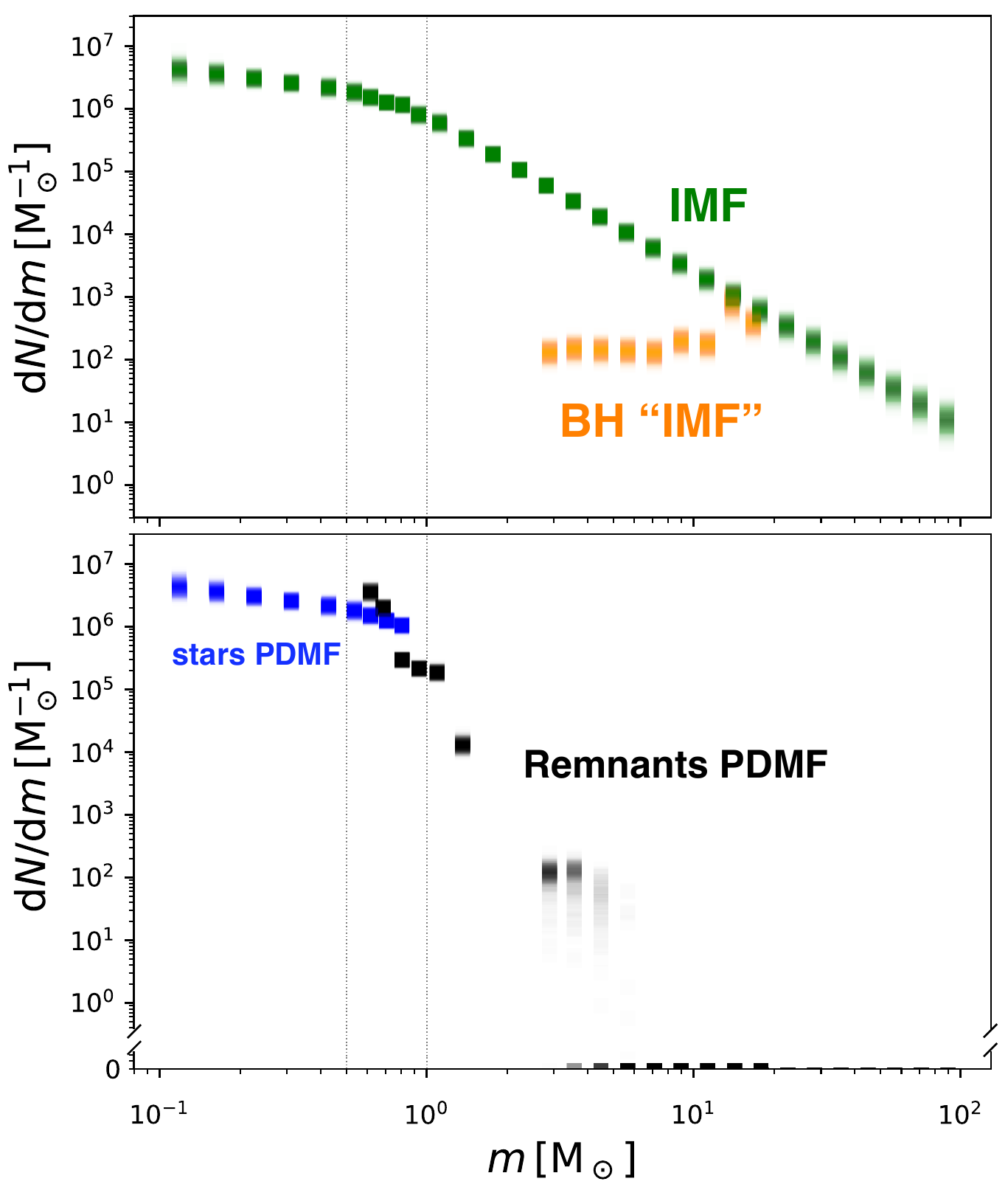}
\caption{Inferred IMF (upper panel, green), initial mass function of BHs (i.e. before any dynamical ejections; upper panel, orange), and present-day mass function of stars (lower panel, blue), and stellar remnants (lower panel, black) based on 200 random samples from the posterior distribution of our model fit to 47 Tuc. We show the binned mass functions that are used as input to {\sc limepy} models. The darker shades are where the different samples pile up, while lighter shades indicate less likely values.}
\label{PDMF_IMF}
\end{figure}

We have seen in the previous section that the global stellar mass function of 47 Tuc is significantly shallower at low masses than that of a `canonical' \citep{2001MNRAS.322..231K} IMF. Because of the long half-mass relaxation time of 47~Tuc ($\log(t_{\rm rh}/{\rm yr})\simeq9.7$), and its relatively circular orbit \citep{2018A&A...616A..12G}, it is unlikely that the { long-term} preferential escape of low-mass stars as the result of equipartition has played an important role. Given our current knowledge of the cluster's orbit, 47 Tuc is not expected to have lost a significant amount of its initial mass  \citep[beyond mass loss from stellar evolution; e.g.][]{TheBalb17}. In fact, dynamical Monte Carlo of models of 47 Tuc require a flat (low-mass) IMF as initial conditions to reproduce a flat present-day mass function \citep{2011MNRAS.410.2698G}.  Other GCs with similar $t_{\rm rh}$ tend to have steeper low-mass mass functions  \citep[corresponding to $\alpha_1$ indices of $\sim1$,][their figure~3]{2017MNRAS.471.3668S}, making 47 Tuc a clear outlier. Fig.~\ref{PDMF_IMF} illustrates the inferred present-day global mass function of stars and remnants from our best-fitting multimass model of 47 Tuc, along with the corresponding initial mass function. 

From integrated light studies of GCs in M31 and the Milky Way, \citet{Strader2011} and \citet{Kimmig2015} found an anti-correlation between the $V-$band mass-to-light ratio ($\Upsilon_V$) and metallicity ([Fe/H]). Because SSP models predict a correlation between $\Upsilon_V$ and [Fe/H], these findings can be interpreted as metal-rich GCs having bottom-light MFs (i.e. fewer dwarfs) and/or top-light MFs (i.e. fewer remnants). However, light-weighted velocity dispersions and half-light radii of mass segregated clusters are sensitive to [Fe/H] dependent biases \citep{2012MNRAS.427..167S}, which can give rise to the observed anti-correlation between the dynamical $\Upsilon_V$ and [Fe/H] even if all GCs formed with the same canonical IMF \citep{2015MNRAS.448L..94S}. The latter authors also showed that the dynamically inferred $\Upsilon_V$ for a canonical MF is similar to that of a bottom-light MF, making $\Upsilon_V$ a particularly poor proxy of the MF at low masses in mass segregated GCs.

Hence resolved GCs are needed to make a solid statement about the MF slope at high [Fe/H]. 47 Tuc is among the most metal-rich Milky Way GCs at a relatively close distance for which kinematics and star counts are available. Our findings suggest that indeed the IMF of this cluster { could have been} both bottom light and (somewhat) top light compared to the canonical IMF.  The universality of the IMF is an ongoing debate, and it can not be concluded yet whether 47~Tuc is a genuine threat to the universality hypothesis or can be reconciled with other GCs once scatter in the outcome of IMF sampling is considered. But it is interesting to note that a putative correlation between MF slope index and [Fe/H] is opposite of what is found in extragalactic studies of early-type galaxies. Systematically larger $\Upsilon_V$ have been reported in more massive (i.e. more metal-rich) early-type galaxies based on dwarf-sensitive spectral features \citep[e.g.][]{2012ApJ...760...71C}, kinematics \citep[e.g.][]{2012Natur.484..485C} and strong lensing \citep[e.g.][]{2010ApJ...709.1195T}. A spectral analysis of (integrated-light properties of) 47 Tuc \citep[as in][]{2017ApJ...850L..14V}, similar to what is done for early-type galaxies, would be an interesting step towards understanding this potentially troubling contradiction between IMF variations in GCs and early-type galaxies.

\subsection{Black holes, the IMF, and the early dynamical evolution of 47 Tuc}
\label{ssec:bhs}
{ There is an important caveat to the argument above about a possible bottom-light IMF for 47 Tuc. Even though it is unlikely that the cluster suffered significant preferential escape of low-mass stars during its long-term evolution (given its long half-mass relaxation time and orbit), there remains a possibility that the loss of low-mass stars occurred early on. If 47 Tuc formed with a high enough density, it could have become completely mass segregated (and thus vulnerable to the loss of low-mass) on a short timescale (an extreme initial half-mass density of $10^7$~M$_\odot$/pc$^3$ would imply an initial half-mass relaxation time of $\sim100$~Myr).

While tides at the present-day orbit are too weak to remove low-mass stars, tidal perturbations with giant molecular clouds (GMCs) during the early evolution could have played a role\footnote{ 47 Tuc is an {\it in situ} cluster according to its position in the age-[Fe/H] diagram \citep[e.g.][]{2013MNRAS.436..122L}, and may have formed in the disc. It could have been scattered out of the plane by GMC interactions, losing (low-mass) stars in the process.}, although for a large initial density the timescale for this process would also be very large. It also remains to be seen whether this can lead to a change in the low-mass mass function by $\Delta \alpha\simeq0.8$ and explain the inferred present-day mass function with a standard IMF. From \citet{2003MNRAS.340..227B}, a cluster needs to lose about 85\% of its initial mass by evaporation to achieve $\Delta \alpha\sim0.8$. About 45\% of this mass loss is due to stellar evolution, so the cluster needs to lose about 40\% of its initial mass through evaporation. For a present day mass of $\sim1\times10^6  \ \msun$, this implies an initial mass of  $\sim6.7\times10^6 \ \msun$. { However, we need to be cautious with using this relation between mass loss and $\Delta \alpha$ to other mass loss mechanism, because the way low-mass stars are lost could be different. For example, tidal perturbations remove stars at large radii, while evaporation removes stars with high energies, and this may not lead to the same $\Delta \alpha$ for a given amount of mass loss. }}

{ Interestingly, a very high initial density might explain why 47 Tuc has retained almost no BHs, which is otherwise puzzling given its relatively long relaxation time and large radius \citep[see e.g.][]{2013MNRAS.432.2779B}. Dense initial conditions are needed to dynamically eject most BHs during a Hubble time.{ The Monte Carlo model of 47~Tuc by \citet{2011MNRAS.410.2698G} has an initial half-mass density of $\sim3\times10^4\ \msun\ {\rm pc}^{-3}$, but we recall that in their models the BHs received the same (large) natal velocity kicks as the neutron stars. Using the fast cluster evolution model of \citet{ag2019} and assuming that the BHs receive the same momentum kick as the neutron stars, we find that an initial density of $\sim10^6\ \msun\ {\rm pc}^{-3}$ is required to dynamically eject nearly all BHs (assuming an initial mass in BHs of 4\% of the total initial mass, appropriate for the metallicity of 47~Tuc). } 
 Although speculative, it could be that more metal-rich GCs like 47 Tuc form substantially denser (due to more efficient cooling) and fully segregate soon after formation, which leads to a flattening of their mass function by various disruption processes like interactions with GMCs.
{ A potential complication is that the high initial density required to quickly mass segregate the cluster and dynamically eject the BHs makes the cluster more resilient to tidal perturbations and makes it harder to reach the present-day $r_{\rm h}\simeq8\ $pc.
 }
 In future modelling efforts, understanding the interplay between the evolution of the cluster size, black hole population, and mass function will be key to understand the IMF of 47~Tuc and the evolution of its BH population.}

\section{Conclusions}
\label{conclusion_section}

We presented a self-consistent multimass {\sc limepy} model of 47 Tuc where the mass function and content of stellar remnants were free to vary and constrained by finding the best simultanenous match to a range of observational constraints: number density profile, kinematics (line-of-sight and proper motion dispersion profiles), and stellar mass function at different projected distances from the cluster centre. Not only does our best-fitting model satisfyingly match all these observations, it also correctly predicts the radial distribution of MSPs in 47 Tuc and can accomodate the line-of-sight accelerations of MSPs inferred from their period derivatives. Our main findings can be summarized as follows:

\begin{itemize}
    \item There is no need for an IMBH in 47 Tuc given the current data. Our model, which does not include an IMBH, can successfully explain a range of observables including the line-of-sight accelerations (or upper limits on these) of MSPs. We thus concur with the conclusions of \citet{2019ApJ...875....1M} rebutting the findings of \citet{2017Natur.542..203K}.
    
    \item We constrain the stellar-mass BH population in 47 Tuc to contain a total mass of $430^{+386}_{-301}$~$\msun$, or equivalently $141^{+97}_{-95}$ BHs { (for a mean BH mass of $3.1 \ \msun$)}. This serves as a demonstration of the power of the mass modelling technique presented here to address questions about the BH content of GCs by modelling their present-day properties. These results also provide important constraints on the dynamical evolution history of 47 Tuc { and its BH population}.
    
    \item We infer a relatively shallow present-day global stellar MF and IMF for 47 Tuc ($\alpha_1=0.52^{+0.17}_{-0.16}$ and $\alpha_2=1.35^{+0.25}_{-0.23}$ compared to the steeper values of $\alpha_1=1.3$ and $\alpha_2=2.3$ for a Kroupa IMF), in keeping with previous results from evolutionary dynamical models of this cluster \citep[e.g.][]{2011MNRAS.410.2698G}. Given our current knowledge of the orbit and predictions for the { long-term} mass-loss history of 47 Tuc, it is not expected to have lost a significant amount of its initial mass (beyond mass loss from stellar evolution), and preferential loss of low-mass stars due to { long-term} dynamical evolution in the tidal field of the Milky Way should have been minimal. The shallow global present-day mass function inferred for main-sequence stars therefore hints at a { possible} dearth of low-mass stars and a bottom-light IMF. { However, we cannot exclude that a loss of low-mass stars occurred early on in the evolution of 47 Tuc. Interestingly, the high initial density that would be required in this case may be linked to the fact that the cluster has retained almost no BHs despite its relatively long relaxation time and large radius at the present day. We emphasize that understanding the interplay between the evolution of the cluster size, black hole population, and mass function will be key to understand the IMF of 47~Tuc and the evolution of its BH population.}
    \item At higher masses, our inferred IMF slope is close to Salpeter ($\alpha_3=-2.49\pm0.08$ compared to $\alpha=2.35$ for Salpeter), although this slope might be steeper if we included the contribution of heavy binaries. In any case, the method presented here, which relies on modelling the dynamical effect of otherwise unseen dark remnants (mainly white dwarfs), provides a promising new way to probe the IMF at high masses in GCs and to address questions about possible variations with environment.
\end{itemize}

\section*{Acknowledgements}
{ We thank the anonymous referee for their report. We also thank Mirek Giersz, Sebastian Kamann, and Holger Baumgardt for useful comments and discussions on a previous version of the manuscript {and {\L}ukasz Wyrzykowski} for insightful discussions on microlensing of BHs.} We would like to thank Laura Watkins for providing the {\it HST} proper motion dispersion profiles for 47 Tuc from \citet{2015ApJ...812..149W}, Jeremy Heyl for sharing the {\it HST} proper motion data from \citet{2017ApJ...850..186H}, and Antonio Sollima for mass function measurements based on the ACS Survey data for 47 Tuc as performed in \citet{2017MNRAS.471.3668S}. VHB thanks S\'ebastien Fabbro for computing support and acknowledges support from the NRC-Canada Plaskett Fellowship. MG, EB and MP acknowledge financial support from the Royal Society (University Research Fellowship) and the
European Research Council (ERC StG-335936, CLUSTERS). JS acknowledges support from the Packard Foundation. Support for Program number HST-GO-14203.002 was provided by NASA through a grant from the Space Telescope Science Institute. We finally thank the International Space Science Institute (ISSI, Bern, CH) for welcoming the meetings of the Team 407 ``Globular Clusters in the {\it Gaia} Era" (team leaders VHB \& MG), during which part of this work was performed.

\bibliographystyle{mnras} 
\bibliography{47Tucbib}

\begin{thebibliography}{}
\makeatletter
\relax
\def\mn@urlcharsother{\let\do\@makeother \do\$\do\&\do\#\do\^\do\_\do\%\do\~}
\def\mn@doi{\begingroup\mn@urlcharsother \@ifnextchar [ {\mn@doi@}
  {\mn@doi@[]}}
\def\mn@doi@[#1]#2{\def\@tempa{#1}\ifx\@tempa\@empty \href
  {http://dx.doi.org/#2} {doi:#2}\else \href {http://dx.doi.org/#2} {#1}\fi
  \endgroup}
\def\mn@eprint#1#2{\mn@eprint@#1:#2::\@nil}
\def\mn@eprint@arXiv#1{\href {http://arxiv.org/abs/#1} {{\tt arXiv:#1}}}
\def\mn@eprint@dblp#1{\href {http://dblp.uni-trier.de/rec/bibtex/#1.xml}
  {dblp:#1}}
\def\mn@eprint@#1:#2:#3:#4\@nil{\def\@tempa {#1}\def\@tempb {#2}\def\@tempc
  {#3}\ifx \@tempc \@empty \let \@tempc \@tempb \let \@tempb \@tempa \fi \ifx
  \@tempb \@empty \def\@tempb {arXiv}\fi \@ifundefined
  {mn@eprint@\@tempb}{\@tempb:\@tempc}{\expandafter \expandafter \csname
  mn@eprint@\@tempb\endcsname \expandafter{\@tempc}}}

\bibitem[\protect\citeauthoryear{{Abbott} et~al.,}{{Abbott}
  et~al.}{2016a}]{2016PhRvL.116f1102A}
{Abbott} B.~P.,  et~al., 2016a, \mn@doi [Physical Review Letters]
  {10.1103/PhysRevLett.116.061102}, \href
  {http://adsabs.harvard.edu/abs/2016PhRvL.116f1102A} {116, 061102}

\bibitem[\protect\citeauthoryear{{Abbott} et~al.,}{{Abbott}
  et~al.}{2016b}]{2016ApJ...818L..22A}
{Abbott} B.~P.,  et~al., 2016b, \mn@doi [\apjl] {10.3847/2041-8205/818/2/L22},
  \href {http://adsabs.harvard.edu/abs/2016ApJ...818L..22A} {818, L22}

\bibitem[\protect\citeauthoryear{{Antonini} \& {Gieles}}{{Antonini} \&
  {Gieles}}{2019}]{ag2019}
{Antonini} F.,  {Gieles} M.,  2019, arXiv:1906.11855, \href
  {https://ui.adsabs.harvard.edu/abs/2019arXiv190611855A} {p. arXiv:1906.11855}

\bibitem[\protect\citeauthoryear{{Antonini}, {Gieles}  \&
  {Gualandris}}{{Antonini} et~al.}{2019}]{2019MNRAS.486.5008A}
{Antonini} F.,  {Gieles} M.,   {Gualandris} A.,  2019, \mn@doi [\mnras]
  {10.1093/mnras/stz1149}, \href
  {https://ui.adsabs.harvard.edu/abs/2019MNRAS.486.5008A} {486, 5008}

\bibitem[\protect\citeauthoryear{{Arca Sedda}, {Askar}  \& {Giersz}}{{Arca
  Sedda} et~al.}{2018}]{2018MNRAS.479.4652A}
{Arca Sedda} M.,  {Askar} A.,   {Giersz} M.,  2018, \mn@doi [\mnras]
  {10.1093/mnras/sty1859}, \href
  {https://ui.adsabs.harvard.edu/abs/2018MNRAS.479.4652A} {479, 4652}

\bibitem[\protect\citeauthoryear{{Askar}, {Arca Sedda}  \& {Giersz}}{{Askar}
  et~al.}{2018}]{2018MNRAS.478.1844A}
{Askar} A.,  {Arca Sedda} M.,   {Giersz} M.,  2018, \mn@doi [\mnras]
  {10.1093/mnras/sty1186}, \href
  {http://adsabs.harvard.edu/abs/2018MNRAS.478.1844A} {478, 1844}

\bibitem[\protect\citeauthoryear{{Bahcall} \& {Wolf}}{{Bahcall} \&
  {Wolf}}{1976}]{1976ApJ...209..214B}
{Bahcall} J.~N.,  {Wolf} R.~A.,  1976, \mn@doi [\apj] {10.1086/154711}, \href
  {http://adsabs.harvard.edu/abs/1976ApJ...209..214B} {209, 214}

\bibitem[\protect\citeauthoryear{{Bahramian} et~al.,}{{Bahramian}
  et~al.}{2017}]{2017MNRAS.467.2199B}
{Bahramian} A.,  et~al., 2017, \mn@doi [\mnras] {10.1093/mnras/stx166}, \href
  {https://ui.adsabs.harvard.edu/abs/2017MNRAS.467.2199B} {467, 2199}

\bibitem[\protect\citeauthoryear{{Balbinot} \& {Gieles}}{{Balbinot} \&
  {Gieles}}{2018}]{TheBalb17}
{Balbinot} E.,  {Gieles} M.,  2018, \mn@doi [\mnras] {10.1093/mnras/stx2708},
  \href {https://ui.adsabs.harvard.edu/abs/2018MNRAS.474.2479B} {474, 2479}

\bibitem[\protect\citeauthoryear{{Baumgardt}}{{Baumgardt}}{2017}]{2017MNRAS.464.2174B}
{Baumgardt} H.,  2017, \mn@doi [\mnras] {10.1093/mnras/stw2488}, \href
  {http://adsabs.harvard.edu/abs/2017MNRAS.464.2174B} {464, 2174}

\bibitem[\protect\citeauthoryear{{Baumgardt} \& {Hilker}}{{Baumgardt} \&
  {Hilker}}{2018}]{2018MNRAS.478.1520B}
{Baumgardt} H.,  {Hilker} M.,  2018, \mn@doi [\mnras] {10.1093/mnras/sty1057},
  \href {http://adsabs.harvard.edu/abs/2018MNRAS.478.1520B} {478, 1520}

\bibitem[\protect\citeauthoryear{{Baumgardt} \& {Makino}}{{Baumgardt} \&
  {Makino}}{2003}]{2003MNRAS.340..227B}
{Baumgardt} H.,  {Makino} J.,  2003, \mn@doi [\mnras]
  {10.1046/j.1365-8711.2003.06286.x}, \href
  {https://ui.adsabs.harvard.edu/abs/2003MNRAS.340..227B} {340, 227}

\bibitem[\protect\citeauthoryear{{Baumgardt} et~al.,}{{Baumgardt}
  et~al.}{2019}]{Baumgardt2019}
{Baumgardt} H.,  et~al., 2019, \mn@doi [\mnras] {10.1093/mnras/stz2060}, \href
  {https://ui.adsabs.harvard.edu/abs/2019MNRAS.tmp.1999B} {p.~1999}

\bibitem[\protect\citeauthoryear{{Belczynski}, {Kalogera}, {Rasio}, {Taam},
  {Zezas}, {Bulik}, {Maccarone}  \& {Ivanova}}{{Belczynski}
  et~al.}{2008}]{2008ApJS..174..223B}
{Belczynski} K.,  {Kalogera} V.,  {Rasio} F.~A.,  {Taam} R.~E.,  {Zezas} A.,
  {Bulik} T.,  {Maccarone} T.~J.,   {Ivanova} N.,  2008, \mn@doi [\apjs]
  {10.1086/521026}, \href
  {https://ui.adsabs.harvard.edu/abs/2008ApJS..174..223B} {174, 223}

\bibitem[\protect\citeauthoryear{{Belczynski}, {Wiktorowicz}, {Fryer}, {Holz}
  \& {Kalogera}}{{Belczynski} et~al.}{2012}]{2012ApJ...757...91B}
{Belczynski} K.,  {Wiktorowicz} G.,  {Fryer} C.~L.,  {Holz} D.~E.,   {Kalogera}
  V.,  2012, \mn@doi [\apj] {10.1088/0004-637X/757/1/91}, \href
  {https://ui.adsabs.harvard.edu/abs/2012ApJ...757...91B} {757, 91}

\bibitem[\protect\citeauthoryear{{Bellini} et~al.,}{{Bellini}
  et~al.}{2014}]{2014ApJ...797..115B}
{Bellini} A.,  et~al., 2014, \mn@doi [\apj] {10.1088/0004-637X/797/2/115},
  \href {http://adsabs.harvard.edu/abs/2014ApJ...797..115B} {797, 115}

\bibitem[\protect\citeauthoryear{{Bellini}, {Bianchini}, {Varri}, {Anderson},
  {Piotto}, {van der Marel}, {Vesperini}  \& {Watkins}}{{Bellini}
  et~al.}{2017}]{2017ApJ...844..167B}
{Bellini} A.,  {Bianchini} P.,  {Varri} A.~L.,  {Anderson} J.,  {Piotto} G.,
  {van der Marel} R.~P.,  {Vesperini} E.,   {Watkins} L.~L.,  2017, \mn@doi
  [\apj] {10.3847/1538-4357/aa7c5f}, \href
  {http://adsabs.harvard.edu/abs/2017ApJ...844..167B} {844, 167}

\bibitem[\protect\citeauthoryear{{Bianchini}, {Norris}, {van de Ven}  \&
  {Schinnerer}}{{Bianchini} et~al.}{2015}]{2015MNRAS.453..365B}
{Bianchini} P.,  {Norris} M.~A.,  {van de Ven} G.,   {Schinnerer} E.,  2015,
  \mn@doi [\mnras] {10.1093/mnras/stv1651}, \href
  {http://adsabs.harvard.edu/abs/2015MNRAS.453..365B} {453, 365}

\bibitem[\protect\citeauthoryear{{Bianchini}, {van de Ven}, {Norris},
  {Schinnerer}  \& {Varri}}{{Bianchini} et~al.}{2016}]{2016MNRAS.458.3644B}
{Bianchini} P.,  {van de Ven} G.,  {Norris} M.~A.,  {Schinnerer} E.,   {Varri}
  A.~L.,  2016, \mn@doi [\mnras] {10.1093/mnras/stw552}, \href
  {http://adsabs.harvard.edu/abs/2016MNRAS.458.3644B} {458, 3644}

\bibitem[\protect\citeauthoryear{{Breen} \& {Heggie}}{{Breen} \&
  {Heggie}}{2013}]{2013MNRAS.432.2779B}
{Breen} P.~G.,  {Heggie} D.~C.,  2013, \mn@doi [\mnras] {10.1093/mnras/stt628},
  \href {https://ui.adsabs.harvard.edu/abs/2013MNRAS.432.2779B} {432, 2779}

\bibitem[\protect\citeauthoryear{{Cappellari} et~al.,}{{Cappellari}
  et~al.}{2012}]{2012Natur.484..485C}
{Cappellari} M.,  et~al., 2012, \mn@doi [\nat] {10.1038/nature10972}, \href
  {https://ui.adsabs.harvard.edu/abs/2012Natur.484..485C} {484, 485}

\bibitem[\protect\citeauthoryear{{Carretta} et~al.,}{{Carretta}
  et~al.}{2009}]{2009A&A...505..117C}
{Carretta} E.,  et~al., 2009, \mn@doi [\aap] {10.1051/0004-6361/200912096},
  \href {http://adsabs.harvard.edu/abs/2009A%26A...505..117C} {505, 117}

\bibitem[\protect\citeauthoryear{{Chen}, {Richer}, {Caiazzo}  \& {Heyl}}{{Chen}
  et~al.}{2018}]{2018ApJ...867..132C}
{Chen} S.,  {Richer} H.,  {Caiazzo} I.,   {Heyl} J.,  2018, \mn@doi [\apj]
  {10.3847/1538-4357/aae089}, \href
  {http://adsabs.harvard.edu/abs/2018ApJ...867..132C} {867, 132}

\bibitem[\protect\citeauthoryear{{Church}, {Strader}, {Davies}  \&
  {Bobrick}}{{Church} et~al.}{2017}]{2017ApJ...851L...4C}
{Church} R.~P.,  {Strader} J.,  {Davies} M.~B.,   {Bobrick} A.,  2017, \mn@doi
  [\apjl] {10.3847/2041-8213/aa9aeb}, \href
  {https://ui.adsabs.harvard.edu/abs/2017ApJ...851L...4C} {851, L4}

\bibitem[\protect\citeauthoryear{{Claydon}, {Gieles}  \& {Zocchi}}{{Claydon}
  et~al.}{2017}]{2017MNRAS.466.3937C}
{Claydon} I.,  {Gieles} M.,   {Zocchi} A.,  2017, \mn@doi [\mnras]
  {10.1093/mnras/stw3309}, \href
  {http://adsabs.harvard.edu/abs/2017MNRAS.466.3937C} {466, 3937}

\bibitem[\protect\citeauthoryear{{Claydon}, {Gieles}, {Varri}, {Heggie}  \&
  {Zocchi}}{{Claydon} et~al.}{2019}]{2019MNRAS.487..147C}
{Claydon} I.,  {Gieles} M.,  {Varri} A.~L.,  {Heggie} D.~C.,   {Zocchi} A.,
  2019, \mn@doi [\mnras] {10.1093/mnras/stz1109}, \href
  {https://ui.adsabs.harvard.edu/abs/2019MNRAS.487..147C} {487, 147}

\bibitem[\protect\citeauthoryear{{Conroy} \& {van Dokkum}}{{Conroy} \& {van
  Dokkum}}{2012}]{2012ApJ...760...71C}
{Conroy} C.,  {van Dokkum} P.~G.,  2012, \mn@doi [\apj]
  {10.1088/0004-637X/760/1/71}, \href
  {https://ui.adsabs.harvard.edu/abs/2012ApJ...760...71C} {760, 71}

\bibitem[\protect\citeauthoryear{{Da Costa}}{{Da
  Costa}}{2016}]{2016MNRAS.455..199D}
{Da Costa} G.~S.,  2016, \mn@doi [\mnras] {10.1093/mnras/stv2315}, \href
  {http://adsabs.harvard.edu/abs/2016MNRAS.455..199D} {455, 199}

\bibitem[\protect\citeauthoryear{{Daniel}, {Heggie}  \& {Varri}}{{Daniel}
  et~al.}{2017}]{2017MNRAS.468.1453D}
{Daniel} K.~J.,  {Heggie} D.~C.,   {Varri} A.~L.,  2017, \mn@doi [\mnras]
  {10.1093/mnras/stx571}, \href
  {https://ui.adsabs.harvard.edu/abs/2017MNRAS.468.1453D} {468, 1453}

\bibitem[\protect\citeauthoryear{{Dotter}}{{Dotter}}{2016}]{2016ApJS..222....8D}
{Dotter} A.,  2016, \mn@doi [\apjs] {10.3847/0067-0049/222/1/8}, \href
  {https://ui.adsabs.harvard.edu/abs/2016ApJS..222....8D} {222, 8}

\bibitem[\protect\citeauthoryear{{Dotter}, {Chaboyer}, {Jevremovi{\'c}},
  {Baron}, {Ferguson}, {Sarajedini}  \& {Anderson}}{{Dotter}
  et~al.}{2007}]{2007AJ....134..376D}
{Dotter} A.,  {Chaboyer} B.,  {Jevremovi{\'c}} D.,  {Baron} E.,  {Ferguson}
  J.~W.,  {Sarajedini} A.,   {Anderson} J.,  2007, \mn@doi [\aj]
  {10.1086/517915}, \href
  {https://ui.adsabs.harvard.edu/abs/2007AJ....134..376D} {134, 376}

\bibitem[\protect\citeauthoryear{{Dotter} et~al.,}{{Dotter}
  et~al.}{2010}]{2010ApJ...708..698D}
{Dotter} A.,  et~al., 2010, \mn@doi [\apj] {10.1088/0004-637X/708/1/698}, \href
  {http://adsabs.harvard.edu/abs/2010ApJ...708..698D} {708, 698}

\bibitem[\protect\citeauthoryear{{Ferrarese} \& {Merritt}}{{Ferrarese} \&
  {Merritt}}{2000}]{2000ApJ...539L...9F}
{Ferrarese} L.,  {Merritt} D.,  2000, \mn@doi [\apjl] {10.1086/312838}, \href
  {http://adsabs.harvard.edu/abs/2000ApJ...539L...9F} {539, L9}

\bibitem[\protect\citeauthoryear{{Foreman-Mackey}, {Hogg}, {Lang}  \&
  {Goodman}}{{Foreman-Mackey} et~al.}{2013}]{2013PASP..125..306F}
{Foreman-Mackey} D.,  {Hogg} D.~W.,  {Lang} D.,   {Goodman} J.,  2013, \mn@doi
  [\pasp] {10.1086/670067}, \href
  {http://adsabs.harvard.edu/abs/2013PASP..125..306F} {125, 306}

\bibitem[\protect\citeauthoryear{{Freire} \& {Ridolfi}}{{Freire} \&
  {Ridolfi}}{2018}]{2018MNRAS.476.4794F}
{Freire} P.~C.~C.,  {Ridolfi} A.,  2018, \mn@doi [\mnras]
  {10.1093/mnras/sty524}, \href
  {http://adsabs.harvard.edu/abs/2018MNRAS.476.4794F} {476, 4794}

\bibitem[\protect\citeauthoryear{{Freire} et~al.,}{{Freire}
  et~al.}{2017}]{2017MNRAS.471..857F}
{Freire} P.~C.~C.,  et~al., 2017, \mn@doi [\mnras] {10.1093/mnras/stx1533},
  \href {http://adsabs.harvard.edu/abs/2017MNRAS.471..857F} {471, 857}

\bibitem[\protect\citeauthoryear{{Fukushige} \& {Heggie}}{{Fukushige} \&
  {Heggie}}{2000}]{2000MNRAS.318..753F}
{Fukushige} T.,  {Heggie} D.~C.,  2000, \mn@doi [\mnras]
  {10.1046/j.1365-8711.2000.03811.x}, \href
  {http://adsabs.harvard.edu/abs/2000MNRAS.318..753F} {318, 753}

\bibitem[\protect\citeauthoryear{{Gaia Collaboration} et~al.,}{{Gaia
  Collaboration} et~al.}{2018}]{2018A&A...616A..12G}
{Gaia Collaboration} et~al., 2018, \mn@doi [\aap]
  {10.1051/0004-6361/201832698}, \href
  {https://ui.adsabs.harvard.edu/abs/2018A&A...616A..12G} {616, A12}

\bibitem[\protect\citeauthoryear{{Gebhardt}, {Pryor}, {Williams}  \&
  {Hesser}}{{Gebhardt} et~al.}{1995}]{1995AJ....110.1699G}
{Gebhardt} K.,  {Pryor} C.,  {Williams} T.~B.,   {Hesser} J.~E.,  1995, \mn@doi
  [\aj] {10.1086/117642}, \href
  {http://adsabs.harvard.edu/abs/1995AJ....110.1699G} {110, 1699}

\bibitem[\protect\citeauthoryear{{Gebhardt} et~al.,}{{Gebhardt}
  et~al.}{2000}]{2000ApJ...539L..13G}
{Gebhardt} K.,  et~al., 2000, \mn@doi [\apjl] {10.1086/312840}, \href
  {http://adsabs.harvard.edu/abs/2000ApJ...539L..13G} {539, L13}

\bibitem[\protect\citeauthoryear{{Gieles} \& {Zocchi}}{{Gieles} \&
  {Zocchi}}{2015}]{2015MNRAS.454..576G}
{Gieles} M.,  {Zocchi} A.,  2015, \mn@doi [\mnras] {10.1093/mnras/stv1848},
  \href {http://adsabs.harvard.edu/abs/2015MNRAS.454..576G} {454, 576}

\bibitem[\protect\citeauthoryear{{Gieles}, {Balbinot}, {Yaaqib},
  {H{\'e}nault-Brunet}, {Zocchi}, {Peuten}  \& {Jonker}}{{Gieles}
  et~al.}{2018a}]{2018MNRAS.473.4832G}
{Gieles} M.,  {Balbinot} E.,  {Yaaqib} R.~I.~S.~M.,  {H{\'e}nault-Brunet} V.,
  {Zocchi} A.,  {Peuten} M.,   {Jonker} P.~G.,  2018a, \mn@doi [\mnras]
  {10.1093/mnras/stx2694}, \href
  {http://adsabs.harvard.edu/abs/2018MNRAS.473.4832G} {473, 4832}

\bibitem[\protect\citeauthoryear{{Gieles} et~al.,}{{Gieles}
  et~al.}{2018b}]{2018MNRAS.478.2461G}
{Gieles} M.,  et~al., 2018b, \mn@doi [\mnras] {10.1093/mnras/sty1059}, \href
  {https://ui.adsabs.harvard.edu/abs/2018MNRAS.478.2461G} {478, 2461}

\bibitem[\protect\citeauthoryear{{Giersz} \& {Heggie}}{{Giersz} \&
  {Heggie}}{2011}]{2011MNRAS.410.2698G}
{Giersz} M.,  {Heggie} D.~C.,  2011, \mn@doi [\mnras]
  {10.1111/j.1365-2966.2010.17648.x}, \href
  {http://adsabs.harvard.edu/abs/2011MNRAS.410.2698G} {410, 2698}

\bibitem[\protect\citeauthoryear{{Giersz}, {Leigh}, {Hypki}, {L{\"u}tzgendorf}
  \& {Askar}}{{Giersz} et~al.}{2015}]{2015MNRAS.454.3150G}
{Giersz} M.,  {Leigh} N.,  {Hypki} A.,  {L{\"u}tzgendorf} N.,   {Askar} A.,
  2015, \mn@doi [\mnras] {10.1093/mnras/stv2162}, \href
  {http://adsabs.harvard.edu/abs/2015MNRAS.454.3150G} {454, 3150}

\bibitem[\protect\citeauthoryear{{Giesers} et~al.,}{{Giesers}
  et~al.}{2018}]{2018MNRAS.475L..15G}
{Giesers} B.,  et~al., 2018, \mn@doi [\mnras] {10.1093/mnrasl/slx203}, \href
  {http://adsabs.harvard.edu/abs/2018MNRAS.475L..15G} {475, L15}

\bibitem[\protect\citeauthoryear{{Goldsbury}, {Richer}, {Anderson}, {Dotter},
  {Sarajedini}  \& {Woodley}}{{Goldsbury} et~al.}{2010}]{2010AJ....140.1830G}
{Goldsbury} R.,  {Richer} H.~B.,  {Anderson} J.,  {Dotter} A.,  {Sarajedini}
  A.,   {Woodley} K.,  2010, \mn@doi [\aj] {10.1088/0004-6256/140/6/1830},
  \href {http://adsabs.harvard.edu/abs/2010AJ....140.1830G} {140, 1830}

\bibitem[\protect\citeauthoryear{{Gratton} et~al.,}{{Gratton}
  et~al.}{2013}]{2013A&A...549A..41G}
{Gratton} R.~G.,  et~al., 2013, \mn@doi [\aap] {10.1051/0004-6361/201219976},
  \href {http://adsabs.harvard.edu/abs/2013A%26A...549A..41G} {549, A41}

\bibitem[\protect\citeauthoryear{{Gunn} \& {Griffin}}{{Gunn} \&
  {Griffin}}{1979}]{1979AJ.....84..752G}
{Gunn} J.~E.,  {Griffin} R.~F.,  1979, \mn@doi [\aj] {10.1086/112477}, \href
  {http://adsabs.harvard.edu/abs/1979AJ.....84..752G} {84, 752}

\bibitem[\protect\citeauthoryear{{Hansen} et~al.,}{{Hansen}
  et~al.}{2013}]{2013Natur.500...51H}
{Hansen} B.~M.~S.,  et~al., 2013, \mn@doi [\nat] {10.1038/nature12334}, \href
  {http://adsabs.harvard.edu/abs/2013Natur.500...51H} {500, 51}

\bibitem[\protect\citeauthoryear{{H{\'e}nault-Brunet}, {Gieles}, {Sollima},
  {Watkins}, {Zocchi}, {Claydon}, {Pancino}  \&
  {Baumgardt}}{{H{\'e}nault-Brunet} et~al.}{2019}]{2019MNRAS.483.1400H}
{H{\'e}nault-Brunet} V.,  {Gieles} M.,  {Sollima} A.,  {Watkins} L.~L.,
  {Zocchi} A.,  {Claydon} I.,  {Pancino} E.,   {Baumgardt} H.,  2019, \mn@doi
  [\mnras] {10.1093/mnras/sty3187}, \href
  {https://ui.adsabs.harvard.edu/abs/2019MNRAS.483.1400H} {483, 1400}

\bibitem[\protect\citeauthoryear{{Heyl}, {Caiazzo}, {Richer}, {Anderson},
  {Kalirai}  \& {Parada}}{{Heyl} et~al.}{2017}]{2017ApJ...850..186H}
{Heyl} J.,  {Caiazzo} I.,  {Richer} H.,  {Anderson} J.,  {Kalirai} J.,
  {Parada} J.,  2017, \mn@doi [\apj] {10.3847/1538-4357/aa974f}, \href
  {http://adsabs.harvard.edu/abs/2017ApJ...850..186H} {850, 186}

\bibitem[\protect\citeauthoryear{{Illingworth} \& {King}}{{Illingworth} \&
  {King}}{1977}]{1977ApJ...218L.109I}
{Illingworth} G.,  {King} I.~R.,  1977, \mn@doi [\apjl] {10.1086/182586}, \href
  {http://adsabs.harvard.edu/abs/1977ApJ...218L.109I} {218, L109}

\bibitem[\protect\citeauthoryear{{Kamann} et~al.,}{{Kamann}
  et~al.}{2018}]{2018MNRAS.473.5591K}
{Kamann} S.,  et~al., 2018, \mn@doi [\mnras] {10.1093/mnras/stx2719}, \href
  {http://adsabs.harvard.edu/abs/2018MNRAS.473.5591K} {473, 5591}

\bibitem[\protect\citeauthoryear{{Kimmig}, {Seth}, {Ivans}, {Strader},
  {Caldwell}, {Anderton}  \& {Gregersen}}{{Kimmig} et~al.}{2015}]{Kimmig2015}
{Kimmig} B.,  {Seth} A.,  {Ivans} I.~I.,  {Strader} J.,  {Caldwell} N.,
  {Anderton} T.,   {Gregersen} D.,  2015, \mn@doi [\aj]
  {10.1088/0004-6256/149/2/53}, \href
  {http://adsabs.harvard.edu/abs/2015AJ....149...53K} {149, 53}

\bibitem[\protect\citeauthoryear{{King}}{{King}}{1966}]{1966AJ.....71...64K}
{King} I.~R.,  1966, \mn@doi [\aj] {10.1086/109857}, \href
  {http://adsabs.harvard.edu/abs/1966AJ.....71...64K} {71, 64}

\bibitem[\protect\citeauthoryear{{K{\i}z{\i}ltan}, {Baumgardt}  \&
  {Loeb}}{{K{\i}z{\i}ltan} et~al.}{2017}]{2017Natur.542..203K}
{K{\i}z{\i}ltan} B.,  {Baumgardt} H.,   {Loeb} A.,  2017, \mn@doi [\nat]
  {10.1038/nature21361}, \href
  {http://adsabs.harvard.edu/abs/2017Natur.542..203K} {542, 203}

\bibitem[\protect\citeauthoryear{{Kremer}, {Chatterjee}, {Rodriguez}  \&
  {Rasio}}{{Kremer} et~al.}{2018a}]{2018ApJ...852...29K}
{Kremer} K.,  {Chatterjee} S.,  {Rodriguez} C.~L.,   {Rasio} F.~A.,  2018a,
  \mn@doi [\apj] {10.3847/1538-4357/aa99df}, \href
  {http://adsabs.harvard.edu/abs/2018ApJ...852...29K} {852, 29}

\bibitem[\protect\citeauthoryear{{Kremer}, {Ye}, {Chatterjee}, {Rodriguez}  \&
  {Rasio}}{{Kremer} et~al.}{2018b}]{2018ApJ...855L..15K}
{Kremer} K.,  {Ye} C.~S.,  {Chatterjee} S.,  {Rodriguez} C.~L.,   {Rasio}
  F.~A.,  2018b, \mn@doi [\apjl] {10.3847/2041-8213/aab26c}, \href
  {http://adsabs.harvard.edu/abs/2018ApJ...855L..15K} {855, L15}

\bibitem[\protect\citeauthoryear{{Kremer}, {Chatterjee}, {Ye}, {Rodriguez}  \&
  {Rasio}}{{Kremer} et~al.}{2019}]{2019ApJ...871...38K}
{Kremer} K.,  {Chatterjee} S.,  {Ye} C.~S.,  {Rodriguez} C.~L.,   {Rasio}
  F.~A.,  2019, \mn@doi [\apj] {10.3847/1538-4357/aaf646}, \href
  {http://adsabs.harvard.edu/abs/2019ApJ...871...38K} {871, 38}

\bibitem[\protect\citeauthoryear{{Kroupa}}{{Kroupa}}{2001}]{2001MNRAS.322..231K}
{Kroupa} P.,  2001, \mn@doi [\mnras] {10.1046/j.1365-8711.2001.04022.x}, \href
  {http://adsabs.harvard.edu/abs/2001MNRAS.322..231K} {322, 231}

\bibitem[\protect\citeauthoryear{{Kunder} et~al.,}{{Kunder}
  et~al.}{2017}]{2017AJ....153...75K}
{Kunder} A.,  et~al., 2017, \mn@doi [\aj] {10.3847/1538-3881/153/2/75}, \href
  {http://adsabs.harvard.edu/abs/2017AJ....153...75K} {153, 75}

\bibitem[\protect\citeauthoryear{{Lane}, {Kiss}, {Lewis}, {Ibata}, {Siebert},
  {Bedding}, {Sz{\'e}kely}  \& {Szab{\'o}}}{{Lane}
  et~al.}{2011}]{2011A&A...530A..31L}
{Lane} R.~R.,  {Kiss} L.~L.,  {Lewis} G.~F.,  {Ibata} R.~A.,  {Siebert} A.,
  {Bedding} T.~R.,  {Sz{\'e}kely} P.,   {Szab{\'o}} G.~M.,  2011, \mn@doi
  [\aap] {10.1051/0004-6361/201116660}, \href
  {http://adsabs.harvard.edu/abs/2011A%26A...530A..31L} {530, A31}

\bibitem[\protect\citeauthoryear{{Lanzoni} et~al.,}{{Lanzoni}
  et~al.}{2013}]{2013ApJ...769..107L}
{Lanzoni} B.,  et~al., 2013, \mn@doi [\apj] {10.1088/0004-637X/769/2/107},
  \href {http://adsabs.harvard.edu/abs/2013ApJ...769..107L} {769, 107}

\bibitem[\protect\citeauthoryear{{Leaman}, {VandenBerg}  \& {Mendel}}{{Leaman}
  et~al.}{2013}]{2013MNRAS.436..122L}
{Leaman} R.,  {VandenBerg} D.~A.,   {Mendel} J.~T.,  2013, \mn@doi [\mnras]
  {10.1093/mnras/stt1540}, \href
  {https://ui.adsabs.harvard.edu/abs/2013MNRAS.436..122L} {436, 122}

\bibitem[\protect\citeauthoryear{{L{\"u}tzgendorf}, {Kissler-Patig}, {Noyola},
  {Jalali}, {de Zeeuw}, {Gebhardt}  \& {Baumgardt}}{{L{\"u}tzgendorf}
  et~al.}{2011}]{2011A&A...533A..36L}
{L{\"u}tzgendorf} N.,  {Kissler-Patig} M.,  {Noyola} E.,  {Jalali} B.,  {de
  Zeeuw} P.~T.,  {Gebhardt} K.,   {Baumgardt} H.,  2011, \mn@doi [\aap]
  {10.1051/0004-6361/201116618}, \href
  {http://adsabs.harvard.edu/abs/2011A%26A...533A..36L} {533, A36}

\bibitem[\protect\citeauthoryear{{L{\"u}tzgendorf} et~al.,}{{L{\"u}tzgendorf}
  et~al.}{2012}]{2012A&A...543A..82L}
{L{\"u}tzgendorf} N.,  et~al., 2012, \mn@doi [\aap]
  {10.1051/0004-6361/201219062}, \href
  {http://adsabs.harvard.edu/abs/2012A%26A...543A..82L} {543, A82}

\bibitem[\protect\citeauthoryear{{L{\"u}tzgendorf}, {Baumgardt}  \&
  {Kruijssen}}{{L{\"u}tzgendorf} et~al.}{2013}]{2013A&A...558A.117L}
{L{\"u}tzgendorf} N.,  {Baumgardt} H.,   {Kruijssen} J.~M.~D.,  2013, \mn@doi
  [\aap] {10.1051/0004-6361/201321927}, \href
  {http://adsabs.harvard.edu/abs/2013A%26A...558A.117L} {558, A117}

\bibitem[\protect\citeauthoryear{{Mackey}, {Wilkinson}, {Davies}  \&
  {Gilmore}}{{Mackey} et~al.}{2008}]{2008MNRAS.386...65M}
{Mackey} A.~D.,  {Wilkinson} M.~I.,  {Davies} M.~B.,   {Gilmore} G.~F.,  2008,
  \mn@doi [\mnras] {10.1111/j.1365-2966.2008.13052.x}, \href
  {http://adsabs.harvard.edu/abs/2008MNRAS.386...65M} {386, 65}

\bibitem[\protect\citeauthoryear{{Mann} et~al.,}{{Mann}
  et~al.}{2019}]{2019ApJ...875....1M}
{Mann} C.~R.,  et~al., 2019, \mn@doi [\apj] {10.3847/1538-4357/ab0e6d}, \href
  {https://ui.adsabs.harvard.edu/abs/2019ApJ...875....1M} {875, 1}

\bibitem[\protect\citeauthoryear{{Mayor} et~al.,}{{Mayor}
  et~al.}{1983}]{1983A&AS...54..495M}
{Mayor} M.,  et~al., 1983, \aaps, \href
  {http://adsabs.harvard.edu/abs/1983A%26AS...54..495M} {54, 495}

\bibitem[\protect\citeauthoryear{{McLaughlin} \& {van der Marel}}{{McLaughlin}
  \& {van der Marel}}{2005}]{2005ApJS..161..304M}
{McLaughlin} D.~E.,  {van der Marel} R.~P.,  2005, \mn@doi [\apjs]
  {10.1086/497429}, \href {http://adsabs.harvard.edu/abs/2005ApJS..161..304M}
  {161, 304}

\bibitem[\protect\citeauthoryear{{McLaughlin}, {Anderson}, {Meylan},
  {Gebhardt}, {Pryor}, {Minniti}  \& {Phinney}}{{McLaughlin}
  et~al.}{2006}]{2006ApJS..166..249M}
{McLaughlin} D.~E.,  {Anderson} J.,  {Meylan} G.,  {Gebhardt} K.,  {Pryor} C.,
  {Minniti} D.,   {Phinney} S.,  2006, \mn@doi [\apjs] {10.1086/505692}, \href
  {https://ui.adsabs.harvard.edu/abs/2006ApJS..166..249M} {166, 249}

\bibitem[\protect\citeauthoryear{{Merritt}, {Piatek}, {Portegies Zwart}  \&
  {Hemsendorf}}{{Merritt} et~al.}{2004}]{2004ApJ...608L..25M}
{Merritt} D.,  {Piatek} S.,  {Portegies Zwart} S.,   {Hemsendorf} M.,  2004,
  \mn@doi [\apjl] {10.1086/422252}, \href
  {http://adsabs.harvard.edu/abs/2004ApJ...608L..25M} {608, L25}

\bibitem[\protect\citeauthoryear{{Meylan}, {Dubath}  \& {Mayor}}{{Meylan}
  et~al.}{1991}]{1991ApJ...383..587M}
{Meylan} G.,  {Dubath} P.,   {Mayor} M.,  1991, \mn@doi [\apj]
  {10.1086/170816}, \href {http://adsabs.harvard.edu/abs/1991ApJ...383..587M}
  {383, 587}

\bibitem[\protect\citeauthoryear{{Michie}}{{Michie}}{1963}]{1963MNRAS.125..127M}
{Michie} R.~W.,  1963, \mn@doi [\mnras] {10.1093/mnras/125.2.127}, \href
  {http://adsabs.harvard.edu/abs/1963MNRAS.125..127M} {125, 127}

\bibitem[\protect\citeauthoryear{{Miller-Jones} et~al.,}{{Miller-Jones}
  et~al.}{2015}]{2015MNRAS.453.3918M}
{Miller-Jones} J.~C.~A.,  et~al., 2015, \mn@doi [\mnras]
  {10.1093/mnras/stv1869}, \href
  {http://adsabs.harvard.edu/abs/2015MNRAS.453.3918M} {453, 3918}

\bibitem[\protect\citeauthoryear{{Milone} et~al.,}{{Milone}
  et~al.}{2012}]{2012A&A...540A..16M}
{Milone} A.~P.,  et~al., 2012, \mn@doi [\aap] {10.1051/0004-6361/201016384},
  \href {https://ui.adsabs.harvard.edu/abs/2012A&A...540A..16M} {540, A16}

\bibitem[\protect\citeauthoryear{{Morscher}, {Pattabiraman}, {Rodriguez},
  {Rasio}  \& {Umbreit}}{{Morscher} et~al.}{2015}]{2015ApJ...800....9M}
{Morscher} M.,  {Pattabiraman} B.,  {Rodriguez} C.,  {Rasio} F.~A.,   {Umbreit}
  S.,  2015, \mn@doi [\apj] {10.1088/0004-637X/800/1/9}, \href
  {https://ui.adsabs.harvard.edu/abs/2015ApJ...800....9M} {800, 9}

\bibitem[\protect\citeauthoryear{{Newell}, {Da Costa}  \& {Norris}}{{Newell}
  et~al.}{1976}]{1976ApJ...208L..55N}
{Newell} B.,  {Da Costa} G.~S.,   {Norris} J.,  1976, \mn@doi [\apjl]
  {10.1086/182232}, \href {http://adsabs.harvard.edu/abs/1976ApJ...208L..55N}
  {208, L55}

\bibitem[\protect\citeauthoryear{{Noyola}, {Gebhardt}  \& {Bergmann}}{{Noyola}
  et~al.}{2008}]{2008ApJ...676.1008N}
{Noyola} E.,  {Gebhardt} K.,   {Bergmann} M.,  2008, \mn@doi [\apj]
  {10.1086/529002}, \href {http://adsabs.harvard.edu/abs/2008ApJ...676.1008N}
  {676, 1008}

\bibitem[\protect\citeauthoryear{{Parada}, {Richer}, {Heyl}, {Kalirai}  \&
  {Goldsbury}}{{Parada} et~al.}{2016}]{2016ApJ...826...88P}
{Parada} J.,  {Richer} H.,  {Heyl} J.,  {Kalirai} J.,   {Goldsbury} R.,  2016,
  \mn@doi [\apj] {10.3847/0004-637X/826/1/88}, \href
  {https://ui.adsabs.harvard.edu/abs/2016ApJ...826...88P} {826, 88}

\bibitem[\protect\citeauthoryear{{Perera} et~al.,}{{Perera}
  et~al.}{2017}]{2017MNRAS.468.2114P}
{Perera} B.~B.~P.,  et~al., 2017, \mn@doi [\mnras] {10.1093/mnras/stx501},
  \href {http://adsabs.harvard.edu/abs/2017MNRAS.468.2114P} {468, 2114}

\bibitem[\protect\citeauthoryear{{Peuten}, {Zocchi}, {Gieles}, {Gualandris}  \&
  {H{\'e}nault-Brunet}}{{Peuten} et~al.}{2016}]{2016MNRAS.462.2333P}
{Peuten} M.,  {Zocchi} A.,  {Gieles} M.,  {Gualandris} A.,
  {H{\'e}nault-Brunet} V.,  2016, \mn@doi [\mnras] {10.1093/mnras/stw1726},
  \href {http://adsabs.harvard.edu/abs/2016MNRAS.462.2333P} {462, 2333}

\bibitem[\protect\citeauthoryear{{Peuten}, {Zocchi}, {Gieles}  \&
  {H{\'e}nault-Brunet}}{{Peuten} et~al.}{2017}]{2017MNRAS.470.2736P}
{Peuten} M.,  {Zocchi} A.,  {Gieles} M.,   {H{\'e}nault-Brunet} V.,  2017,
  \mn@doi [\mnras] {10.1093/mnras/stx1311}, \href
  {http://adsabs.harvard.edu/abs/2017MNRAS.470.2736P} {470, 2736}

\bibitem[\protect\citeauthoryear{{Pfahl}, {Rappaport}  \&
  {Podsiadlowski}}{{Pfahl} et~al.}{2002}]{2002ApJ...573..283P}
{Pfahl} E.,  {Rappaport} S.,   {Podsiadlowski} P.,  2002, \mn@doi [\apj]
  {10.1086/340494}, \href
  {https://ui.adsabs.harvard.edu/abs/2002ApJ...573..283P} {573, 283}

\bibitem[\protect\citeauthoryear{{Phinney}}{{Phinney}}{1993}]{1993ASPC...50..141P}
{Phinney} E.~S.,  1993, in {Djorgovski} S.~G.,  {Meylan} G.,  eds,
  Astronomical Society of the Pacific Conference Series Vol. 50, Structure and
  Dynamics of Globular Clusters. p.~141

\bibitem[\protect\citeauthoryear{{Portegies Zwart} \& {McMillan}}{{Portegies
  Zwart} \& {McMillan}}{2000}]{2000ApJ...528L..17P}
{Portegies Zwart} S.~F.,  {McMillan} S.~L.~W.,  2000, \mn@doi [\apjl]
  {10.1086/312422}, \href {http://adsabs.harvard.edu/abs/2000ApJ...528L..17P}
  {528, L17}

\bibitem[\protect\citeauthoryear{{Portegies Zwart}, {Baumgardt}, {Hut},
  {Makino}  \& {McMillan}}{{Portegies Zwart}
  et~al.}{2004}]{2004Natur.428..724P}
{Portegies Zwart} S.~F.,  {Baumgardt} H.,  {Hut} P.,  {Makino} J.,   {McMillan}
  S.~L.~W.,  2004, \mn@doi [\nat] {10.1038/nature02448}, \href
  {http://adsabs.harvard.edu/abs/2004Natur.428..724P} {428, 724}

\bibitem[\protect\citeauthoryear{{Prager}, {Ransom}, {Freire}, {Hessels},
  {Stairs}, {Arras}  \& {Cadelano}}{{Prager}
  et~al.}{2017}]{2017ApJ...845..148P}
{Prager} B.~J.,  {Ransom} S.~M.,  {Freire} P.~C.~C.,  {Hessels} J.~W.~T.,
  {Stairs} I.~H.,  {Arras} P.,   {Cadelano} M.,  2017, \mn@doi [\apj]
  {10.3847/1538-4357/aa7ed7}, \href
  {http://adsabs.harvard.edu/abs/2017ApJ...845..148P} {845, 148}

\bibitem[\protect\citeauthoryear{{Ridolfi} et~al.,}{{Ridolfi}
  et~al.}{2016}]{2016MNRAS.462.2918R}
{Ridolfi} A.,  et~al., 2016, \mn@doi [\mnras] {10.1093/mnras/stw1850}, \href
  {http://adsabs.harvard.edu/abs/2016MNRAS.462.2918R} {462, 2918}

\bibitem[\protect\citeauthoryear{{Rodriguez}, {Chatterjee}  \&
  {Rasio}}{{Rodriguez} et~al.}{2016}]{2016PhRvD..93h4029R}
{Rodriguez} C.~L.,  {Chatterjee} S.,   {Rasio} F.~A.,  2016, \mn@doi [\prd]
  {10.1103/PhysRevD.93.084029}, \href
  {https://ui.adsabs.harvard.edu/\#abs/2016PhRvD..93h4029R} {93, 084029}

\bibitem[\protect\citeauthoryear{{Salpeter}}{{Salpeter}}{1955}]{1955ApJ...121..161S}
{Salpeter} E.~E.,  1955, \mn@doi [\apj] {10.1086/145971}, \href
  {https://ui.adsabs.harvard.edu/abs/1955ApJ...121..161S} {121, 161}

\bibitem[\protect\citeauthoryear{{Sarajedini} et~al.,}{{Sarajedini}
  et~al.}{2007}]{2007AJ....133.1658S}
{Sarajedini} A.,  et~al., 2007, \mn@doi [\aj] {10.1086/511979}, \href
  {http://adsabs.harvard.edu/abs/2007AJ....133.1658S} {133, 1658}

\bibitem[\protect\citeauthoryear{{Shanahan} \& {Gieles}}{{Shanahan} \&
  {Gieles}}{2015}]{2015MNRAS.448L..94S}
{Shanahan} R.~L.,  {Gieles} M.,  2015, \mn@doi [\mnras]
  {10.1093/mnrasl/slu205}, \href
  {http://adsabs.harvard.edu/abs/2015MNRAS.448L..94S} {448, L94}

\bibitem[\protect\citeauthoryear{{Sigurdsson} \& {Hernquist}}{{Sigurdsson} \&
  {Hernquist}}{1993}]{1993Natur.364..423S}
{Sigurdsson} S.,  {Hernquist} L.,  1993, \mn@doi [\nat] {10.1038/364423a0},
  \href {https://ui.adsabs.harvard.edu/abs/1993Natur.364..423S} {364, 423}

\bibitem[\protect\citeauthoryear{{Sippel}, {Hurley}, {Madrid}  \&
  {Harris}}{{Sippel} et~al.}{2012}]{2012MNRAS.427..167S}
{Sippel} A.~C.,  {Hurley} J.~R.,  {Madrid} J.~P.,   {Harris} W.~E.,  2012,
  \mn@doi [\mnras] {10.1111/j.1365-2966.2012.21969.x}, \href
  {https://ui.adsabs.harvard.edu/abs/2012MNRAS.427..167S} {427, 167}

\bibitem[\protect\citeauthoryear{{Sollima} \& {Baumgardt}}{{Sollima} \&
  {Baumgardt}}{2017}]{2017MNRAS.471.3668S}
{Sollima} A.,  {Baumgardt} H.,  2017, \mn@doi [\mnras] {10.1093/mnras/stx1856},
  \href {http://adsabs.harvard.edu/abs/2017MNRAS.471.3668S} {471, 3668}

\bibitem[\protect\citeauthoryear{{Sollima}, {Baumgardt}, {Zocchi}, {Balbinot},
  {Gieles}, {H{\'e}nault-Brunet}  \& {Varri}}{{Sollima}
  et~al.}{2015}]{2015MNRAS.451.2185S}
{Sollima} A.,  {Baumgardt} H.,  {Zocchi} A.,  {Balbinot} E.,  {Gieles} M.,
  {H{\'e}nault-Brunet} V.,   {Varri} A.~L.,  2015, \mn@doi [\mnras]
  {10.1093/mnras/stv1079}, \href
  {http://adsabs.harvard.edu/abs/2015MNRAS.451.2185S} {451, 2185}

\bibitem[\protect\citeauthoryear{{Strader}, {Caldwell}  \& {Seth}}{{Strader}
  et~al.}{2011}]{Strader2011}
{Strader} J.,  {Caldwell} N.,   {Seth} A.~C.,  2011, \mn@doi [\aj]
  {10.1088/0004-6256/142/1/8}, \href
  {http://adsabs.harvard.edu/abs/2011AJ....142....8S} {142, 8}

\bibitem[\protect\citeauthoryear{{Strader}, {Chomiuk}, {Maccarone},
  {Miller-Jones}  \& {Seth}}{{Strader} et~al.}{2012a}]{2012Natur.490...71S}
{Strader} J.,  {Chomiuk} L.,  {Maccarone} T.~J.,  {Miller-Jones} J.~C.~A.,
  {Seth} A.~C.,  2012a, \mn@doi [\nat] {10.1038/nature11490}, \href
  {http://adsabs.harvard.edu/abs/2012Natur.490...71S} {490, 71}

\bibitem[\protect\citeauthoryear{{Strader}, {Chomiuk}, {Maccarone},
  {Miller-Jones}, {Seth}, {Heinke}  \& {Sivakoff}}{{Strader}
  et~al.}{2012b}]{2012ApJ...750L..27S}
{Strader} J.,  {Chomiuk} L.,  {Maccarone} T.~J.,  {Miller-Jones} J.~C.~A.,
  {Seth} A.~C.,  {Heinke} C.~O.,   {Sivakoff} G.~R.,  2012b, \mn@doi [\apjl]
  {10.1088/2041-8205/750/2/L27}, \href
  {http://adsabs.harvard.edu/abs/2012ApJ...750L..27S} {750, L27}

\bibitem[\protect\citeauthoryear{{Tiongco}, {Vesperini}  \& {Varri}}{{Tiongco}
  et~al.}{2016}]{2016MNRAS.455.3693T}
{Tiongco} M.~A.,  {Vesperini} E.,   {Varri} A.~L.,  2016, \mn@doi [\mnras]
  {10.1093/mnras/stv2574}, \href
  {http://adsabs.harvard.edu/abs/2016MNRAS.455.3693T} {455, 3693}

\bibitem[\protect\citeauthoryear{{Trager}, {King}  \& {Djorgovski}}{{Trager}
  et~al.}{1995}]{1995AJ....109..218T}
{Trager} S.~C.,  {King} I.~R.,   {Djorgovski} S.,  1995, \mn@doi [\aj]
  {10.1086/117268}, \href {http://adsabs.harvard.edu/abs/1995AJ....109..218T}
  {109, 218}

\bibitem[\protect\citeauthoryear{{Tremou} et~al.,}{{Tremou}
  et~al.}{2018}]{2018arXiv180600259T}
{Tremou} E.,  et~al., 2018, preprint, \href
  {http://adsabs.harvard.edu/abs/2018arXiv180600259T} {} (\mn@eprint {arXiv}
  {1806.00259})

\bibitem[\protect\citeauthoryear{{Treu}, {Auger}, {Koopmans}, {Gavazzi},
  {Marshall}  \& {Bolton}}{{Treu} et~al.}{2010}]{2010ApJ...709.1195T}
{Treu} T.,  {Auger} M.~W.,  {Koopmans} L. V.~E.,  {Gavazzi} R.,  {Marshall}
  P.~J.,   {Bolton} A.~S.,  2010, \mn@doi [\apj]
  {10.1088/0004-637X/709/2/1195}, \href
  {https://ui.adsabs.harvard.edu/abs/2010ApJ...709.1195T} {709, 1195}

\bibitem[\protect\citeauthoryear{{Tudor} et~al.,}{{Tudor}
  et~al.}{2018}]{2018MNRAS.476.1889T}
{Tudor} V.,  et~al., 2018, \mn@doi [\mnras] {10.1093/mnras/sty284}, \href
  {https://ui.adsabs.harvard.edu/abs/2018MNRAS.476.1889T} {476, 1889}

\bibitem[\protect\citeauthoryear{{Villaume}, {Brodie}, {Conroy}, {Romanowsky}
  \& {van Dokkum}}{{Villaume} et~al.}{2017}]{2017ApJ...850L..14V}
{Villaume} A.,  {Brodie} J.,  {Conroy} C.,  {Romanowsky} A.~J.,   {van Dokkum}
  P.,  2017, \mn@doi [\apj] {10.3847/2041-8213/aa970f}, \href
  {https://ui.adsabs.harvard.edu/abs/2017ApJ...850L..14V} {850, L14}

\bibitem[\protect\citeauthoryear{{Volonteri}}{{Volonteri}}{2010}]{2010A&ARv..18..279V}
{Volonteri} M.,  2010, \mn@doi [\aapr] {10.1007/s00159-010-0029-x}, \href
  {http://adsabs.harvard.edu/abs/2010A%26ARv..18..279V} {18, 279}

\bibitem[\protect\citeauthoryear{{Watkins}, {van der Marel}, {Bellini}  \&
  {Anderson}}{{Watkins} et~al.}{2015}]{2015ApJ...812..149W}
{Watkins} L.~L.,  {van der Marel} R.~P.,  {Bellini} A.,   {Anderson} J.,  2015,
  \mn@doi [\apj] {10.1088/0004-637X/812/2/149}, \href
  {http://adsabs.harvard.edu/abs/2015ApJ...812..149W} {812, 149}

\bibitem[\protect\citeauthoryear{{Weatherford}, {Chatterjee}, {Rodriguez}  \&
  {Rasio}}{{Weatherford} et~al.}{2018}]{2018ApJ...864...13W}
{Weatherford} N.~C.,  {Chatterjee} S.,  {Rodriguez} C.~L.,   {Rasio} F.~A.,
  2018, \mn@doi [\apj] {10.3847/1538-4357/aad63d}, \href
  {http://adsabs.harvard.edu/abs/2018ApJ...864...13W} {864, 13}

\bibitem[\protect\citeauthoryear{{Wilson}}{{Wilson}}{1975}]{1975AJ.....80..175W}
{Wilson} C.~P.,  1975, \mn@doi [\aj] {10.1086/111729}, \href
  {http://adsabs.harvard.edu/abs/1975AJ.....80..175W} {80, 175}

\bibitem[\protect\citeauthoryear{{Woolley}}{{Woolley}}{1954}]{1954MNRAS.114..191W}
{Woolley} R.~V.~D.~R.,  1954, \mnras, \href
  {http://adsabs.harvard.edu/abs/1954MNRAS.114..191W} {114, 191}

\bibitem[\protect\citeauthoryear{{Wyrzykowski} \& {Mandel}}{{Wyrzykowski} \&
  {Mandel}}{2019}]{2019arXiv190407789W}
{Wyrzykowski} {\L}.,  {Mandel} I.,  2019, arXiv:1904.07789, \href
  {https://ui.adsabs.harvard.edu/abs/2019arXiv190407789W} {p. arXiv:1904.07789}

\bibitem[\protect\citeauthoryear{{Wyrzykowski} et~al.,}{{Wyrzykowski}
  et~al.}{2016}]{2016MNRAS.458.3012W}
{Wyrzykowski} {\L}.,  et~al., 2016, \mn@doi [\mnras] {10.1093/mnras/stw426},
  \href {https://ui.adsabs.harvard.edu/abs/2016MNRAS.458.3012W} {458, 3012}

\bibitem[\protect\citeauthoryear{{Zocchi}, {Gieles}, {H{\'e}nault-Brunet}  \&
  {Varri}}{{Zocchi} et~al.}{2016}]{2016MNRAS.462..696Z}
{Zocchi} A.,  {Gieles} M.,  {H{\'e}nault-Brunet} V.,   {Varri} A.~L.,  2016,
  \mn@doi [\mnras] {10.1093/mnras/stw1104}, \href
  {http://adsabs.harvard.edu/abs/2016MNRAS.462..696Z} {462, 696}

\bibitem[\protect\citeauthoryear{{Zocchi}, {Gieles}  \&
  {H{\'e}nault-Brunet}}{{Zocchi} et~al.}{2017}]{2017MNRAS.468.4429Z}
{Zocchi} A.,  {Gieles} M.,   {H{\'e}nault-Brunet} V.,  2017, \mn@doi [\mnras]
  {10.1093/mnras/stx316}, \href
  {http://adsabs.harvard.edu/abs/2017MNRAS.468.4429Z} {468, 4429}

\bibitem[\protect\citeauthoryear{{Zocchi}, {Gieles}  \&
  {H{\'e}nault-Brunet}}{{Zocchi} et~al.}{2019}]{2019MNRAS.482.4713Z}
{Zocchi} A.,  {Gieles} M.,   {H{\'e}nault-Brunet} V.,  2019, \mn@doi [\mnras]
  {10.1093/mnras/sty1508}, \href
  {http://adsabs.harvard.edu/abs/2019MNRAS.482.4713Z} {482, 4713}

\bibitem[\protect\citeauthoryear{{de Boer}, {Gieles}, {Balbinot},
  {H{\'e}nault-Brunet}, {Sollima}, {Watkins}  \& {Claydon}}{{de Boer}
  et~al.}{2019}]{2019MNRAS.tmp..651D}
{de Boer} T.~J.~L.,  {Gieles} M.,  {Balbinot} E.,  {H{\'e}nault-Brunet} V.,
  {Sollima} A.,  {Watkins} L.~L.,   {Claydon} I.,  2019, \mn@doi [\mnras]
  {10.1093/mnras/stz651}, \href
  {http://adsabs.harvard.edu/abs/2019MNRAS.tmp..651D} {}

\bibitem[\protect\citeauthoryear{{de Vita}, {Trenti}, {Bianchini}, {Askar},
  {Giersz}  \& {van de Ven}}{{de Vita} et~al.}{2017}]{2017MNRAS.467.4057D}
{de Vita} R.,  {Trenti} M.,  {Bianchini} P.,  {Askar} A.,  {Giersz} M.,   {van
  de Ven} G.,  2017, \mn@doi [\mnras] {10.1093/mnras/stx325}, \href
  {http://adsabs.harvard.edu/abs/2017MNRAS.467.4057D} {467, 4057}

\bibitem[\protect\citeauthoryear{{den Brok}, {van de Ven}, {van den Bosch}  \&
  {Watkins}}{{den Brok} et~al.}{2014}]{2014MNRAS.438..487D}
{den Brok} M.,  {van de Ven} G.,  {van den Bosch} R.,   {Watkins} L.,  2014,
  \mn@doi [\mnras] {10.1093/mnras/stt2221}, \href
  {http://adsabs.harvard.edu/abs/2014MNRAS.438..487D} {438, 487}

\bibitem[\protect\citeauthoryear{{van der Marel} \& {Anderson}}{{van der Marel}
  \& {Anderson}}{2010}]{2010ApJ...710.1063V}
{van der Marel} R.~P.,  {Anderson} J.,  2010, \mn@doi [\apj]
  {10.1088/0004-637X/710/2/1063}, \href
  {http://adsabs.harvard.edu/abs/2010ApJ...710.1063V} {710, 1063}

\makeatother
\end{thebibliography}

\label{lastpage}

\end{document}